%% file: ms.tex
\documentclass[aps,prl,twocolumn,english,superscriptaddress,showpacs,longbibliography]{revtex4-1}
\usepackage{amsmath}
\usepackage{amssymb,scalerel}
\usepackage{graphicx}
\usepackage[colorlinks,bookmarks=false,citecolor=red,linkcolor=blue,urlcolor=blue]{hyperref}
\usepackage{lipsum, color}
\usepackage{wasysym}

\def\i{{\boldsymbol i}}
\def\j{{\boldsymbol j}}
\def\k{{\boldsymbol k}}

\def\Q{{\boldsymbol Q}}
\def\S{{\boldsymbol S}}

\newcommand{\ve}[1]{\boldsymbol{#1}}

\graphicspath{ {./RESULTS_NEW/} }

\begin{document}
\title{  Breakdown of heavy  quasiparticles  in a  honeycomb Kondo lattice: \\ A  quantum Monte Carlo  study   }
\author{Marcin Raczkowski}
\affiliation{Institut f\"ur Theoretische Physik und Astrophysik, Universit\"at W\"urzburg, 97074 W\"urzburg, Germany}
\author{Bimla Danu}
\affiliation{Institut f\"ur Theoretische Physik und Astrophysik, Universit\"at W\"urzburg, 97074 W\"urzburg, Germany}
\author{Fakher F. Assaad}
\affiliation{Institut f\"ur Theoretische Physik und Astrophysik and W\"urzburg-Dresden Cluster of Excellence ct.qmat, Universit\"at W\"urzburg, 97074 W\"urzburg, Germany}
\date{\today}
\begin{abstract}
We  show  that   for the half-filled  Kondo lattice model  on the  honeycomb lattice   a Kondo breakdown  occurs   
at   small Kondo couplings $J_k$  within   the   magnetically ordered phase.  
Our  conclusions are
based on  auxiliary field  quantum Monte Carlo simulations of  the  so-called  composite  fermion spectral  function.   
Within a  U(1)   gauge  theory  formulation of the Kondo model, it  becomes apparent  
that  a  Higgs   mechanism  dictates the weight of  the resonance   in the  spectral  function.  
For  the    honeycomb lattice  we  observe that for small $J_k$  the quasiparticle pole  gives  way  to  incoherent  
spectral  weight  but it remains well defined for the square lattice. 
Our  result provides  an  explicit  example  where   
the magnetic  transition  and   the breakdown of heavy  quasiparticles  are   detached as  observed  in  
	Yb(Rh$_{0.93}$Co$_{0.07}$)$_2$Si$_2$ [Friedemann et al., Nat. Phys. \textbf{5}, 465 (2009)].
\end{abstract}

\maketitle

Strongly   correlated many  body   systems  are   characterized by  the emergence of new  elementary  excitations.  This  can occur  
through  the   fractionalization of  the electron within a    parton  type  construction --  fractional  quantum Hall  effect~\cite{Jain89}  
or   Luttinger liquids~\cite{Giamarchi} --   or through  the formation  of   a  composite object.  Examples of  the latter    range  
from the understanding of single-hole dynamics in  quantum antiferromagnets~\cite{BERAN1996707,Grusdt18}  to  the emergence  of  the   
electron   in    $\mathbb{Z}_2$  lattice  gauge  theories  in which  the  electron is a bound state of an  orthogonal  fermion   and   
$\mathbb{Z}_2$  matter~\cite{Nandkishore12,Gazit19,Hohenadler18}.

 \begin{figure}[b]
\includegraphics[width=0.5\textwidth]{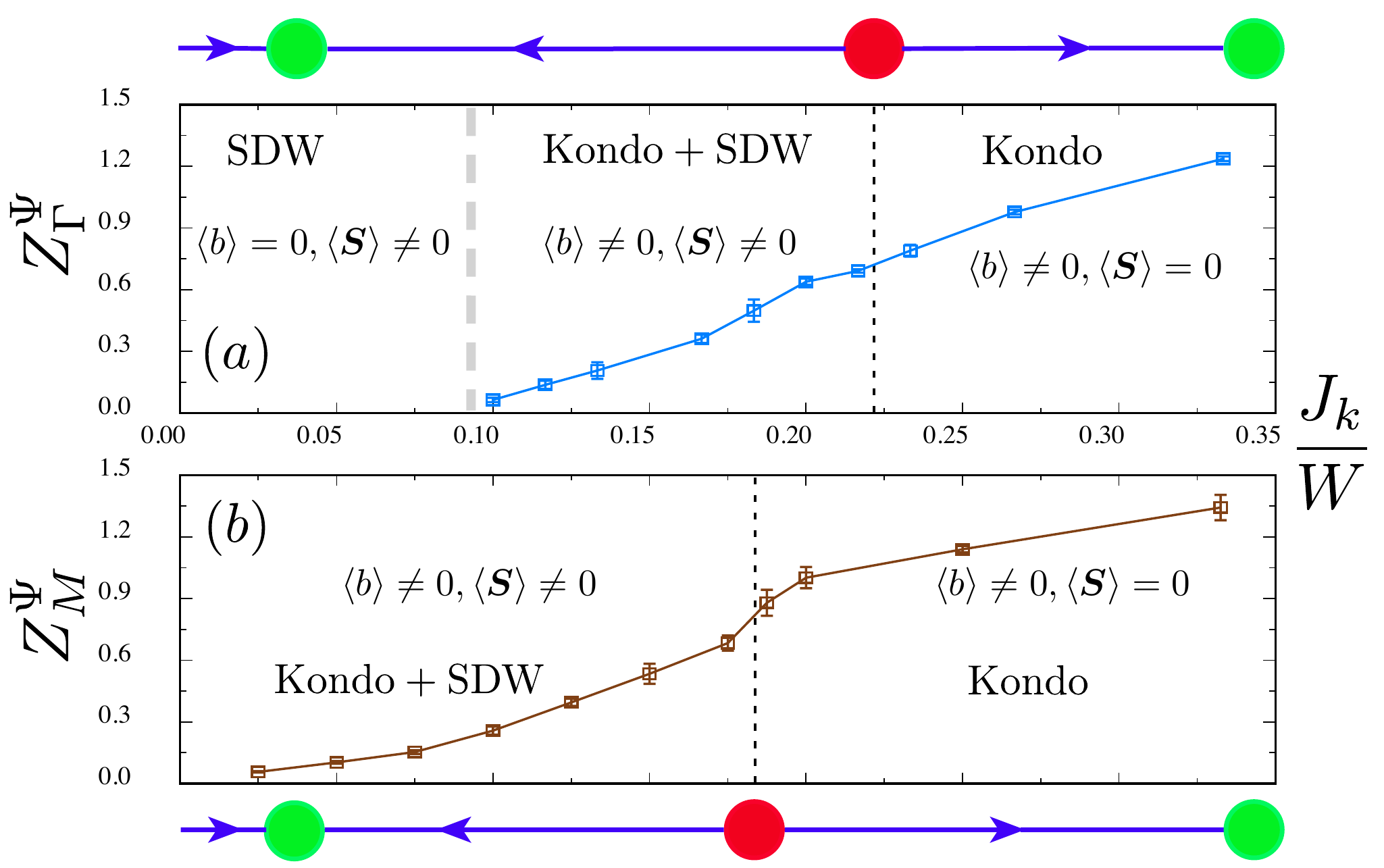}
\caption{Ground-state phase diagram   of  the half-filled Kondo lattice  model on  the honeycomb and square lattices.    
On both  lattices   we  observe a  magnetic order-disorder  transition denoted  by a red circle and  order parameter  corresponding to $ \langle \ve{S} \rangle $.      
For  the honeycomb lattice  (a)  we  observe  a  breakdown  of  the heavy quasiparticle in the spin-density-wave (SDW) phase as indicated 
by the vanishing residue 
$Z^{\psi}_{\ve{k}}  $ of the pole  at the $\Gamma$ point in the composite fermion Green's function.   
For   the  square lattice  (b)  we  observe only  the order-disorder  transition since, down to our lowest value 
$J_k/W=0.025$, $Z^{\psi}_{\ve{k}}$ at the $M=(\pi,\pi)$ point remains finite. 
All the values of $Z^{\psi}_{\ve{k}}$ are  extrapolated to the thermodynamic limit~\cite{SM}.  
We  use  the mean-field  notation,  
$\langle  b  \rangle$,   to  track the magnitude of  the residue. 
}
\label{Fig:phase}
\end{figure}

The  Kondo  effect  is  yet  another  example of  the emergence  of a  composite  fermion  carrying  the  quantum numbers of the  electron. 
Consider  a spin-1/2 magnetic impurity embedded in  Fermi liquid with finite density of states at the Fermi energy.  In the presence of time 
reversal symmetry the Kondo  coupling between the impurity and Fermi liquid is always relevant  and  leads  to the  emergence of a composite  
fermion. It    consists of  the spin-1/2  and  conduction electrons and  becomes itinerant 
thereby  releasing the   $\ln(2)$ entropy.  
 If one replaces the metal by a Dirac liquid with vanishing density of states at the Fermi energy, the Kondo coupling is irrelevant and one 
 will generically observe a transition from an unscreened to screened moment at finite value of the  Kondo coupling~\cite{Withoff90,Fritz04}.   
 This  transition corresponds to the   breakdown   of  the  aforementioned  composite  fermion  \cite{Si01,Coleman01}.     Such  phenomena are not limited to the  
 realm of  impurity  physics~\cite{PhysRevB.95.104402}.   
 Neutron scattering experiments of  metallic Yb$_2$Pt$_2$Pb~\cite{Wu16}     suggest  a  Kondo breakdown phase 
 of  a one-dimensional   spin  chain   embedded in a  three-dimensional metal.    Furthermore,  numerical  evidence  of this  state of matter  
 has  been observed   in models of spin chains on  semimetals~\cite{Danu20}.   In  dense  systems  such as  in 
 YbRh$_2$Si$_2$~\cite{Paschen04,Friedmann10,Prochaska20}  or CeCoIn$_5$~\cite{Maksimovic22},  
 the  notion  of Kondo  breakdown or  orbital Mott  selective  transitions~\cite{Vojta10}  has  deep  implications since   
 the  composite    fermions    drop  out  from the  Luttinger  count.  
 For  systems with an \textit{odd}  number  of    localized  spins per   unit     cell    and  no further   spontaneous symmetry breaking,    
 this  implies  a violation of  the Luttinger sum rule.    Owing to Oshikawa's~\cite{Oshikawa00a}  work   such a  violation  can  be understood 
 if  the   spin system  shows  topological degeneracy     akin of a spin liquid~\cite{Senthil03,Hofmann18}.    
For  an  \textit{even}  number of   spins   per  unit  cell,  such  topological  constraints  do not  hold.  
In this  case,  Kondo breakdown   does  not imply  a   violation of Luttinger's  theorem. 

Since the tight binding model on the honeycomb lattice provides a realization of Dirac electrons,    one  may  ask  the 
question  if and how the aforementioned Kondo breakdown  transition in   the impurity   limit~\cite{Withoff90,Fritz04} 
is  carried over  to  the  dense  case described by the half-filled  Kondo lattice  model.
  In  Ref.~\cite{Yamamoto07} it is  argued  that   the Kondo coupling   is   marginal  in the  weak  coupling limit   thereby  opening  the possibility of  Kondo  breakdown transitions  in magnetically  ordered   metallic  states.  
The  central  result  of  this  Letter   is summarized in Fig.~\ref{Fig:phase}:   Kondo  breakdown  indeed  occurs  within  the  magnetic 
phase  of the honeycomb lattice. In contrast no breakdown is  observed on the   square  lattice.

\textit{U(1) gauge  theory  approach.}
  Since  the Kondo effect  and  concomitant  emergence  of  the composite fermion  is  not    related  to   spontaneous symmetry  breaking, 
  some  care  has  to be  taken  in defining the  onset of these phenomena.  They become  particularly transparent  within a  U(1)   
  gauge  theory  approach  to  the Kondo lattice  model~\cite{Read83,Auerbach86,Saremi07}.  
        The  Kondo lattice model (KLM)  on the  honeycomb lattice reads:
   \begin{equation}
   \hat{H}_{KLM} = \sum_{\ve{i},\ve{j}}    T_{\ve{i},\ve{j}} \ve{\hat{c}}^{\dagger}_{\ve{i}}  \ve{\hat{c}}^{\phantom\dagger}_{\ve{j}}  + 
        \frac{J_k}{2}   \sum_{\ve{i}}    \ve{\hat{c}}^{\dagger}_{\ve{i}} \ve{\sigma} \ve{\hat{c}}^{\phantom\dagger}_{\ve{i}} \cdot \ve{\hat{S}}^{}_{\ve{i}},
\end{equation}
where  $\ve{\hat{c}}^{\dagger}_{\ve{i}}    = \left( \hat{c}^{\dagger}_{\ve{i},\uparrow},  \hat{c}^{\dagger}_{\ve{i},\downarrow}  \right) $  is a
  spinor   where ${\hat c}^{\dagger}_{\i,\sigma}$ creates an electron  in Wannier  state  centered  around lattice site $\ve{i}$ and 
  $z$ component of spin $\sigma = \uparrow,\downarrow$.   $J_k$ is the Kondo exchange coupling between conduction electrons and   spins s=1/2,  
  $\ve{\hat{S}}_{\ve{i}}  $, with $\ve{\sigma}$  being   a vector of Pauli spin matrices.     The  matrix $T_{\ve{i},\ve{j}}$   accounts  for  
  nearest  neighbor hopping  with  amplitude $-t$.  We  adopt  an     Abrikosov  representation of the spin  operator,  
$ \ve{\hat{S}}^{}_{\ve{i}}    =    \frac{1}{2} \ve{\hat{f}}^{\dagger}_{\ve{i}} \ve{\sigma} \ve{\hat{f}}^{\phantom\dagger}_{\ve{i}}  $   with  
$ \ve{\hat{f}}^{\dagger}_{\ve{i}}    = \left( \hat{f}^{\dagger}_{\ve{i},\uparrow},  \hat{f}^{\dagger}_{\ve{i},\downarrow}  \right) $    
and  constraint $ \ve{\hat{f}}^{\dagger}_{\ve{i}} \ve{\hat{f}}^{\phantom\dagger}_{\ve{i}}  = 1  $.     
To   proceed we  use  the   following rewriting of  the  Kondo term  $ -  \frac{J_k}{4}   \left(  \hat{V}^{\dagger}_{\ve{i}}  \hat{V}^{\phantom\dagger}_{\ve{i}}  +  \hat{V}^{\phantom\dagger}_{\ve{i}}  \hat{V}^{\dagger}_{\ve{i}}   \right)   $ with  
$ \hat{V}^{\dagger}_{\ve{i}}   =   \ve{\hat{c}}^{\dagger}_{\ve{i}}  \ve{\hat{f}}^{\phantom\dagger}_{\ve{i}}  $.   
In the  constrained  Hilbert  space,  this rewriting is  exact.  To  formulate  the path  integral,  we  will   work in an unconstrained  
Hilbert  space   and  impose it energetically  with a   Hubbard-$U$  term: 
$H_U =  U \sum_{\ve{i}} \left(  \ve{\hat{f}}^{\dagger}_{\ve{i}}  \ve{\hat{f}}^{\phantom\dagger}_{\ve{i}}  -1 \right)^2 $.   
Importantly the  fermion parity  on the  $f$-orbitals  is   a  constant of  motion  such  that  it is    very efficient to implement in 
numerical simulations.     We  can decouple the Kondo  (Hubbard)  term  with 
a complex (real)   field,  $b_{\ve{i}}(\tau)$,   $a_{0,\ve{i}} (\tau) $  to  obtain  the  following action  in terms of  Grassmann  variables 
$\ve{f}_{\ve{i}}(\tau) $  and $\ve{c}_{\ve{i}}(\tau) $ 
\begin{widetext}
\begin{eqnarray}
\label{eq:KLM_U1S_m}
	S  =   S_0^{c} +   \int_{0}^{\beta}  d \tau  & &  \left\{   \sum_{\ve{i}}  \left[ \frac{2}{J_k}  |  b_{\ve{i}}(\tau)  |^2  +  i    a_{0,\ve{i}}(\tau) + 
	 \ve{f}_{\ve{i}}^{\dagger}(\tau) \left[ \partial_\tau - i  a_{0,\ve{i}}(\tau)  \right] \ve{f}_{\ve{i}}^{\phantom\dagger}(\tau)    
	        + b_{\ve{i}}(\tau) \ve{c}^{\dagger}_{\ve{i}}  \ve{f}^{\phantom\dagger}_{\ve{i}}   +  
	            \overline{b_{\ve{i}}(\tau)}  \ve{f}^{\dagger}_{\ve{i}}  \ve{c}^{\phantom\dagger}_{\ve{i}}   \right]   
	              \right\}  
\end{eqnarray}
\end{widetext} 
with 
$  S_0^{c}= \int_{0}^{\beta} d \tau  \sum_{\ve{i},\ve{j}}     
	         \ve{c}_{\ve{i}}^{\dagger}(\tau) \left[ \partial_\tau \delta_{\ve{i},\ve{j}} +   T_{\ve{i},\ve{j}}  \right] \ve{c}_{\ve{j}}^{\phantom\dagger}(\tau)  $. 
 The  above  corresponds  to the  action in the limit $U \rightarrow \infty $   where  local  U(1)   gauge invariance is  apparent.   
 In  particular  the   canonical transformation,   $\ve{f}_{\ve{i}} (\tau)   \rightarrow   \ve{f}_{\ve{i}} (\tau) e^{i \chi_{\ve{i}}(\tau)} $
amounts  to  redefining the  fields $  a_{0,\ve{i}}(\tau) \rightarrow a_{0,\ve{i}}(\tau)  + \partial_\tau \chi_{\ve{i}}(\tau)  $    and 
$ b_{\ve{i}}(\tau)  \rightarrow      b_{\ve{i}}(\tau) 	 e^{-i \chi_{\ve{i}}(\tau)}  $, such  that the  partition function remains  invariant.    We  are  now in a position to  probe  for  various phases    with  gauge invariant quantities.     
Magnetism, triggered by the RKKY interaction,  corresponds  to a  spontaneous    global  SU(2) spin symmetry   
breaking  and   long  ranged  correlations    of  the order  parameter  
$\ve{\hat{S}}_{\ve{i}} = \frac{1}{2} \hat{\ve{f}}^{\dagger}_{\ve{i}} \ve{\sigma} \hat{\ve{f}}^{\phantom\dagger}_{\ve{i}}  $.   
Clearly   $\ve{\hat{S}}_{\ve{i}} $  carries no  U(1)   charge.  To  define  the Kondo  effect  we  consider  the fermion   field
\begin{equation}
\label{eq:tildef}
	 \ve{\tilde{f}}_{\ve{i}}(\tau)   =   e^{i \varphi_{\ve{i}}(\tau)}  \ve{f}_{\ve{i}}(\tau),   \text{ with }   e^{i \varphi_{\ve{i}}(\tau)}  =  \frac{b_{\ve{i}} (\tau)}
	 {|b_{\ve{i}} (\tau)|}.
\end{equation}
  As   argued in the  Supplemental  Material~\cite{SM}, $\ve{\tilde{f}}_{\ve{i}}(\tau) $  has  the  quantum numbers of  a   physical  fermion:  
  it carries no gauge charge, has an electron charge $e$,  and  spin 1/2.       The  Kondo effect  corresponds  to  the emergence of  this 
  fermion at low  energies  as signaled by a pole (resonance)  in the dense case (single impurity limit)  in the  
  corresponding spectral  function~\cite{Raczkowski18,Danu19}.   
  There is no symmetry   that imposes 
  $\langle   \ve{\tilde{f}}^{\phantom\dagger}_{\ve{i}}(\tau)  \ve{\tilde{f}}^{\dagger}_{\ve{j}}(\tau')   \rangle $  to  vanish  between  
  two  space-time points and  the pole  in the corresponding  spectral function    reflects this  fact.   Furthermore, if  the  ground state  
  turns out  to be  a Fermi liquid, the  Luttinger   volume  will  have  to account  for  the composite  fermion. 
 
The  above  can  be  understood  in terms of a   Higgs~\cite{Fradkin79}  mechanism    in  which  the phase  fluctuations of  
$\varphi_{\ve{i}}(\tau)$    become very slow  such   that   $\varphi_{\ve{i}}(\tau)$    can be  set to  a  constant.    In this  case   
there is  no  distinction  between  $ \ve{\tilde{f}}_{\ve{i}}(\tau)   $  and $\ve{f}_{\ve{i}}(\tau)$   or,  in other  words,   
$\ve{f}_{\ve{i}}(\tau)$   has  lost its gauge  charge  and  has    acquired  a unit   electric charge.  This  Higgs  mechanism  is  
captured  in  mean-field large-$N$  approaches  of  the Kondo  lattice    where  Kondo  screening  corresponds  to  
$ \left< b_{\ve{i}} (\tau)     \right> \neq 0 $~\cite{Burdin00,Morr17}.  

The  above  definition of the fermion field,   $\tilde{\ve{f}}$,  depends  explicitly on the  gauge  field  that is  not  accessible in    
generic  numerical simulations  (e.g.   exact   diagonalization).    However,  reintroducing  amplitude  fluctuations of the  $b$-field,  
we   have
$
	    \ve{\tilde{f}} _{\ve{i}}  \propto   b_{i}(\tau)  \ve{f}_{\ve{i}}(\tau) \propto    
	     \left[ \ve{f}^{\dagger}_{\ve{i}}(\tau) \ve{c}^{\phantom\dagger}_{\ve{i}}(\tau)   \right]  
	     \ve{f}_{\ve{i}}(\tau).
$ 
As  shown in  Ref.~\cite{Danu21}  and in  the large-$N$ limit,  the   right  hand  side of the latter equation is   nothing  but  the  
composite  fermion  field: 
\begin{equation}	
	\ve{\tilde{f}} _{\ve{i}}  \propto   \ve{\psi}_{\ve{i}} =    \ve{S}^{}_{\ve{i}} \cdot   \ve{\sigma}  \ve{c}_{\ve{i}}. 
\end{equation}
We also note  that    $\langle  b^{}_{i} b^{\dagger}_{i}   \rangle   \propto  \langle \hat{V}^{\dagger}_{i} \hat{V}^{}_{i}   \rangle    \propto  \langle  \hat{\ve{c}}^{\dagger}_{i}  \ve{\sigma} \hat{\ve{c}}^{}_{i}   \cdot \hat{\ve{S}}_{i} \rangle$  such  that  
the  local  spin  correlations between  the  conduction electrons  and  impurity  spins    correspond  to  the modulus of   the  boson field.    
If  this  quantity   remains  finite  in the considered parameter  regime,  we will conclude    that an  adequate gauge  field  independent 
representation of $\ve{\tilde{f}} _{\ve{i}} $ is given by the composite  fermion field $\ve{\psi}_{\ve{i}}$~\cite{Costi00,Maltseva09}.  
   For impurity problems the  Green's function of   $ \hat{\ve{\psi}}^{\dagger}_{\i} $ corresponds  to the   $T$-matrix~\cite{Borda07} while   
   $ \hat{\ve{\psi}}^{\dagger}_{\i}  $  itself    corresponds  to the    Schrieffer-Wolff   transformation of the localized electron
  operator  in the  realm of the Anderson model~\cite{Raczkowski18}.

\textit{Method.} 
For   our  simulations  we  use  the  projective (zero-temperature) version of the  Algorithms for Lattice  Fermions (ALF)~\cite{ALF_v2}  
implementation of the   auxiliary  field  quantum Monte Carlo (QMC)  method~\cite{Blankenbecler81,Sugiyama86,White89,Sorella89,Assaad08_rev}. 
For a proper comparison between honeycomb and square lattices,  we set hereafter their respective tight binding bandwidths 
$W=6t$ and $W=8t$ as the energy units.

\input{results_new.tex}

\textit{Summary and conclusions.}  
We have investigated  a Kondo breakdown    defined by  the   \textit{destruction}  of  the composite fermion 
in Eq.~(\ref{eq:tildef}).  
In the  realm  of  the Kondo lattice  considered  here,  this amounts to the loss of a pole  in the  
composite  fermion  Green's  function.  Our  main result,  is  that Kondo  breakdown occurs  in the  magnetic  phase   
of  the half-filled   KLM  on the honeycomb lattice.   This  stands in  stark  contrast  to  our  results on the  
square   lattice  where   down  to  the  lowest  values of  the Kondo coupling, we  observe  no breakdown  of the composite fermion.

Our  results  show  that the  magnetic  transition and  Kondo breakdown   are   detached as  observed  in 
Yb(Rh$_{0.93}$Co$_{0.07}$)$_2$Si$_2$~\cite{Friedmann09}.  
The observed  Kondo breakdown  corresponds   to a  modification of  the  excitation  spectra,  and  does  not  
necessarily   translate into   a  thermodynamic  transition.   This stands   in agreement  with the  Fradkin-Shenker~\cite{Fradkin79}  
phase  diagram  where confined   and  Higgs  phases are  adiabatically  connected.   It  would  be  of   great interest  to   modify  the KLM 
so as  to allow  for  a  deconfined phase    and  probe  the  full  richness of  the  Fradkin-Shenker  phase  diagram  as  suggested 
in Ref.~\cite{Saremi07}.
On the experimental side, we hope that our results will have an impact on the studies aimed at exploring  
quantum impurity problems in graphene in a dense situation~\cite{Fritz_2013,Jiang18,Hwang18}.

\begin{acknowledgments}
We  thank  T.  Grover  and  M. Vojta  for   many  illuminating  conversations and  Zihong Liu  for  work on a  related  project. 
The authors gratefully acknowledge the Gauss Centre for Supercomputing e.V. (www.gauss-centre.eu) for funding this project 
by providing computing time on the GCS Supercomputer SUPERMUC-NG at Leibniz Supercomputing Centre (www.lrz.de) as well as 
through the John von Neumann Institute for Computing (NIC) on the GCS Supercomputer JUWELS  \cite{JUWELS} at the J\"ulich 
Supercomputing Centre (JSC).  
M.R. is funded by the Deutsche Forschungsgemeinschaft (DFG, German Research Foundation), Project No. 332790403.
B.D. thanks the W\"urzburg-Dresden Cluster of Excellence on Complexity and Topology in Quantum Matter ct.qmat (EXC 2147, project-id
390858490) for  financial support.   F.F.A. and  B.D. acknowledge  support from the DFG funded SFB 1170
on Topological and Correlated Electronics at Surfaces and Interfaces.
\end{acknowledgments}

\bibliographystyle{bibstyle}
\bibliography{marcin,fassaad,ref}

\clearpage
\widetext
\renewcommand\theequation{S\arabic{equation}}
\renewcommand\thefigure{S\arabic{figure}}
\setcounter{figure}{0}
\setcounter{section}{0}

\newcounter{myequation}
\makeatletter
\@addtoreset{figure}{section}
\@addtoreset{equation}{myequation}
\makeatother

\stepcounter{myequation}

\begin{center}
 \large{ \bf Supplemental Material for:  \\ Breakdown of heavy  quasiparticles  in a  honeycomb Kondo lattice: \\ A  quantum Monte Carlo  study 
  }  
\end{center}
\section{ Auxiliary  field  QMC  and  U(1)   gauge  theory  of  the Kondo Lattice  model }

To  at best  understand the phases of the Kondo lattice model, 
\begin{eqnarray}
 \hat{H}_{KLM} &= &\sum_{\ve{i},\ve{j}}    T_{\ve{i},\ve{j}} \ve{\hat{c}}^{\dagger}_{\ve{i}}  \ve{\hat{c}}^{\phantom\dagger}_{\ve{j}}  + 
        \frac{J_k}{2}   \sum_{\ve{i}}    \ve{\hat{c}}^{\dagger}_{\ve{i}} \ve{\sigma} \ve{\hat{c}}^{\phantom\dagger}_{\ve{i}} \cdot \ve{\hat{S}}^{}_{\ve{i}},
\label{eqS:KLM}
\end{eqnarray}
where  $\ve{\hat{c}}^{\dagger}_{\ve{i}}    = \left( \hat{c}^{\dagger}_{\ve{i},\uparrow},  \hat{c}^{\dagger}_{\ve{i},\downarrow}  \right) $  
is a      Wannier-state spinor,  we  adopt  an     Abrikosov  representation of the spin  operator,  
$ \ve{\hat{S}}^{}_{\ve{i}}    =    \frac{1}{2} \ve{\hat{f}}^{\dagger}_{\ve{i}} \ve{\sigma} \ve{\hat{f}}^{\phantom\dagger}_{\ve{i}}  $   with  
$ \ve{\hat{f}}^{\dagger}_{\ve{i}}    = \left( \hat{f}^{\dagger}_{\ve{i},\uparrow},  \hat{f}^{\dagger}_{\ve{i},\downarrow}  \right) $    and  constraint
$ \ve{\hat{f}}^{\dagger}_{\ve{i}} \ve{\hat{f}}^{\phantom\dagger}_{\ve{i}}  = 1  $.   In the  constrained    Hilbert  space,  the  identity: 
\begin{equation}
	  \frac{J_k}{2}     \ve{\hat{c}}^{\dagger}_{\ve{i}} \ve{\sigma} \ve{\hat{c}}^{\phantom\dagger}_{\ve{i}} \cdot \ve{\hat{S}}^{}_{\ve{i}}    =  
	    -  \frac{J_k}{4}   \left(  \hat{V}^{\dagger}_{\ve{i}}  \hat{V}^{\phantom\dagger}_{\ve{i}}  +  \hat{V}^{\phantom\dagger}_{\ve{i}}  \hat{V}^{\dagger}_{\ve{i}}   \right)  
\end{equation}
with  $ \hat{V}^{\dagger}_{\ve{i}}   =   \ve{\hat{c}}^{\dagger}_{\ve{i}}  \ve{\hat{f}}^{\phantom\dagger}_{\ve{i}}  $  holds.        
To  proceed  we   relax  the constraint  and  consider  the Hamiltonian
\begin{equation}
	 \hat{H}_{KLM}   =  \lim_{U \rightarrow \infty} \left\{   \sum_{\ve{i},\ve{j}}    T_{\ve{i},\ve{j}} \ve{\hat{c}}^{\dagger}_{\ve{i}}  \ve{\hat{c}}^{\phantom\dagger}_{\ve{j}}         
      -   \frac{J_k}{8}   \sum_{\ve{i}}  \left(  \left[ \hat{V}^{\dagger}_{i}  +  \hat{V}^{\phantom\dagger}_{i}   \right]^2 +  \left[  i \hat{V}^{\dagger}_{i}   - 
          i \hat{V}^{\phantom\dagger}_{i}  \right]^2 \right)
      +   \frac{U}{2}   \sum_{\ve{i}}  \left(   \ve{\hat{f}}^{\dagger}_{\ve{i}}     \ve{\hat{f}}^{\phantom\dagger}_{\ve{i}}   -1  \right)^2 \right\}. 
\end{equation}
Since  the  Hubbard  term  commutes  with  the  Hamiltonian,   the  projection  onto  the   physical   Hilbert  space   occurs  at  
a  rate  set  by    
$  \left< \left(   \ve{\hat{f}}^{\dagger}_{\ve{i}}     \ve{\hat{f}}^{\phantom\dagger}_{\ve{i}}   -1  \right)^2   \right>    \propto  e^{- \beta  U/2 }  $,   where  $\beta$  corresponds to the  inverse  temperature.       Using  the  Trotter   decomposition,   and  Hubbard Stratonovich    
transformation    to    decouple  the   perfect  square terms   we  obtain the  following  form  for the  grand-canonical  partition 
function~\cite{Negele}:
\begin{equation}
	   Z    =   \text{Tr}   e^{ - \beta  \hat{H}_{KLM} }  \propto \int  D \left\{  \ve{f}^{\dagger}_{\ve{i}}(\tau),  \ve{f}^{\phantom\dagger}_{\ve{i}}(\tau), 
	     \ve{c}^{\dagger}_{\ve{i}}(\tau),  \ve{c}^{\phantom\dagger}_{\ve{i}}(\tau),   b_{\ve{i}}(\tau),    a_{0,\ve{i}}(\tau)   \right\}     e^{-S}
\end{equation}
with  
\begin{eqnarray}
\label{eq:KLM_U1S}
	S  =    \int_{0}^{\beta}  d \tau  & &  \left\{   \sum_{\ve{i}}  \left[ \frac{N}{J_k}  |  b_{\ve{i}}(\tau)  |^2  +  \frac{N}{U}  | a_{0,\ve{i}}(\tau) |^2+  i \frac{N}{2}   a_{0,\ve{i}}(\tau) + 
	 \ve{f}_{\ve{i}}^{\dagger}(\tau) \left[ \partial_\tau - i  a_{0,\ve{i}}(\tau)  \right] \ve{f}_{\ve{i}}^{\phantom\dagger}(\tau)    
	        + b_{\ve{i}}(\tau) \ve{c}^{\dagger}_{\ve{i}}  \ve{f}^{\phantom\dagger}_{\ve{i}}   +  
	            \overline{b_{\ve{i}}(\tau)}  \ve{f}^{\dagger}_{\ve{i}}  \ve{c}^{\phantom\dagger}_{\ve{i}}   \right]     \right.  \nonumber  \\
	  &&  \left. 
	       + \sum_{\ve{i},\ve{j}}     
	         \ve{c}_{\ve{i}}^{\dagger}(\tau) \left[ \partial_\tau \delta_{\ve{i},\ve{j}} +   T_{\ve{i},\ve{j}}  \right] \ve{c}_{\ve{j}}^{\phantom\dagger}(\tau)   \right\}.  
\end{eqnarray}
In the  above,  $ a_{0,\ve{i}}(\tau) $  is a  real  field   used  to impose  the constraint, $b_{\ve{i}}(\tau)$  a  complex  field  for  
the Kondo  term,   and   $\ve{c}^{\dagger}_{\ve{i}}$  as  well  as  $  \ve{f}^{\dagger}_{\ve{i}} $  are  spinors  of   Grassmann  variables.   
We  have  also  taken  the liberty of  enhancing  the  spin index  from  $N=2$  to a  general $N$  with    constraint  
$  \ve{f}^{\dagger}_{\ve{i}}  \ve{f}^{\phantom\dagger}_{\ve{i}} =  N/2$~\cite{Raczkowski20,hazra2021luttinger}.    
   The  above  action  is   the starting  point   for auxiliary  field QMC  simulations~\cite{ALF_v2}  as  well as  for the  classification 
   of phases.    For  the QMC  simulations  we   use  the  Gauss-Hermite  quadrature  
to   replace   continuous  fields   by discrete  ones.     The   integration over the  Grassmann  variables yields  the fermion determinant,  
that for    particle-hole symmetric  conduction   electrons   and  even values of $N$  is  positive  semi-definite.   
   The  integration over  the      Hubbard-Stratonovich   fields  is then carried out  
 with Monte  Carlo  importance sampling.    For  details of the  implementation,  we  refer  the reader to Ref.~\cite{ALF_v2}.   
  In particular  for the calculation presented  here,  we  have  used  the  implementation of the Kondo  lattice  model  of the ALF-2.0    
  library.  
 
The  constraint  leads  to a  U(1)  local   gauge invariance.   In  particular,    and  only  in the  $U \rightarrow  \infty $  limit,  
the     canonical transformation 
\begin{equation} 
	\ve{f}	_{\ve{i}} (\tau)   \rightarrow   \ve{f}_{\ve{i}} (\tau) e^{i \chi_{\ve{i}}(\tau)}
\end{equation}
amounts  to  redefining the  fields 
\begin{equation}
	 a_{0,\ve{i}}(\tau) \rightarrow a_{0,\ve{i}}(\tau)  + \partial_\tau \chi_{\ve{i}}(\tau)  \; \;  \text{   and  }  \; \;  b_{\ve{i}}(\tau)  \rightarrow 
b_{\ve{i}}(\tau) 	 e^{-i \chi_{\ve{i}}(\tau)} 
\end{equation}
in the action. 

\section{ Phases of the Kondo Lattice  model }
The  above  action allows us  to   define    precisely  the   two  phases    of  the Kondo lattice model  that are of importance to us in  
the present  article.   The  spin-density-wave (SDW)  phase  is  characterized  by  long ranged  order  in 
\begin{equation}
	 \langle   \frac{1}{2} \ve{\hat{f}}^{\dagger}_{\ve{i}} \ve{\sigma} \ve{\hat{f}}^{\phantom\dagger}_{\ve{i}}     \cdot 
	 \frac{1}{2} \ve{\hat{f}}^{\dagger}_{\ve{j}} \ve{\sigma} \ve{\hat{f}}^{\phantom\dagger}_{\ve{j}}   \rangle   
\end{equation}
and  is  hence characterized by  a    non-vanishing vacuum  expectation value   of   
$ \langle   \frac{1}{2} \ve{\hat{f}}^{\dagger}_{\ve{i}} \ve{\sigma} \ve{\hat{f}}^{\phantom\dagger}_{\ve{i}} \rangle $  in the  thermodynamic 
limit  characteristic of  spontaneous  symmetry  breaking  of  the  global   SU(2)  spin  symmetry. 
Note  that the  above expectation value is  taken  with  respect to the action of  Eq.~(\ref{eq:KLM_U1S}).   

The  Kondo phase   is more  subtle  to  define  since it   is  not  characterized  by a    broken  symmetry.    
Let $b_{\ve{i}} (\tau) =  | b_{\ve{i}} (\tau)|  e^{i \varphi_{\ve{i}}(\tau)} $  and  
\begin{equation}
	   \ve{\tilde{f}}_{\ve{i}}(\tau)   =   e^{i \varphi_{\ve{i}}(\tau)}  \ve{f}_{\ve{i}}(\tau). 
\end{equation} 
$\ve{\tilde{f}}_{\ve{i}}$  is  a physical  fermion operator.  As  mentioned  above,  under a local U(1) gauge  transformation,  $\ve{f}_{\ve{i}}   \rightarrow \ve{f}_{\ve{i}} e^{i \chi_{\ve{i}} (\tau)}$,   $\varphi_{\ve{i}}(\tau)  \rightarrow  \varphi_{\ve{i}}(\tau)  - \chi_{\ve{i}} (\tau) $    such  that   $\ve{\tilde{f}}_{\ve{i}}$   remains invariant. It  hence  carries no   gauge charge.       $ \ve{\tilde{f}}_{\ve{i}} $   
carries  electric  charge.      Consider  the global  $U(1)$   charge    transformation,   $ \hat{T}(\alpha)  =  e^{i  \alpha  \sum_{\ve{i}}   \ve{\hat{c}}^{\dagger}_{\ve{i} }  \ve{\hat{c}}^{\phantom\dagger}_{\ve{i}} }  $.    Since   $ T(\alpha) $  is  a  conserved  quantity,    
and  the   physical  electron  transforms    as 
 $  \hat{T}(\alpha)^{-1} \ve{\hat{c}}^{\phantom\dagger}_{\ve{i}}  \hat{T}(\alpha)   =  e^{i \alpha}\ve{\hat{c}}^{\phantom\dagger}_{\ve{i}} $,    the  phase 
$ \varphi_{\ve{i}}(\tau) $   transforms as  $ \varphi_{\ve{i}}(\tau) \rightarrow \varphi_{\ve{i}}(\tau)  +  \alpha $.
 Hence   $   \ve{\tilde{f}}_{\ve{i}} $   transforms   as the   electron:
 $ \ve{\tilde{f}}_{\ve{i}}  \rightarrow   \ve{\tilde{f}}_{\ve{i}} e^{i\alpha} $.   In  the  heavy  fermion  phase,    the    electron operator 
 $  \ve{\tilde{f}}_{\ve{i}} $   emerges as  a new  particle  excitation that acquires  coherence.  This is  what is   meant in  colloquial  
 terms by  \textit{  the  spins  delocalize  and  participate in the Luttinger  volume}.

Since  in the  heavy  fermions phase $  \ve{\tilde{f}} _{\ve{i}} $    is  the emergent  quasiparticle,  it is natural  to   write the action  
for  this  degree of  freedom.    From  Eq.~(\ref{eq:KLM_U1S})    and in the limit  $U  \rightarrow \infty $, one   readily  obtains: 
\begin{eqnarray}
\label{eq:KLM_U1S_tildef}
	S  =    \int_{0}^{\beta}  d \tau  & &  \left\{   \sum_{\ve{i}}  \left[ \frac{N}{J_k}  |  b_{\ve{i}}(\tau)  |^2  +  i \frac{N}{2}   a_{0,\ve{i}}(\tau) + 
	 \ve{\tilde{f}}_{\ve{i}}^{\dagger}(\tau) \left[ \partial_\tau - i  a_{0,\ve{i}}(\tau)  -i   \partial_{\tau}  \varphi_{\ve{i}}(\tau)  \right] \ve{\tilde{f}}_{\ve{i}}^{\phantom\dagger}(\tau)    
	        + | b_{\ve{i}}(\tau) | \left( \ve{c}^{\dagger}_{\ve{i}}  \ve{\tilde{f}}^{\phantom\dagger}_{\ve{i}}   +  
	             \ve{\tilde{f}}^{\dagger}_{\ve{i}}  \ve{c}^{\phantom\dagger}_{\ve{i}}  \right)  \right]     \right.  \nonumber  \\
	  &&  \left. 
	       + \sum_{\ve{i},\ve{j}}     
	         \ve{c}_{\ve{i}}^{\dagger}(\tau) \left[ \partial_\tau \delta_{\ve{i},\ve{j}} +   T_{\ve{i},\ve{j}}  \right] \ve{c}_{\ve{j}}^{\phantom\dagger}(\tau)   \right\}.  
\end{eqnarray} 
Importantly,   $ \ve{\tilde{f}}^{\phantom\dagger}_{\ve{i}}(\tau)  $  does  not    possess a  local  U(1) gauge   charge  such  that $
	    \langle \ve{\tilde{f}}^{\dagger}_{\ve{i}}  (\tau)  \ve{\tilde{f}}^{\phantom\dagger}_{\ve{j}}  (\tau')  \rangle   $
does  not  vanish  by  symmetry  for   $ ( \ve{i}, \tau )  \ne  ( \ve{j}, \tau' )  $.     In  contrast,  owing  to the local  U(1) symmetry, 
$\langle \ve{f}^{\dagger}_{\ve{i}}  (\tau)  \ve{f}^{\phantom\dagger}_{\ve{j}}  (\tau')  \rangle   = 0  $  
if    $ ( \ve{i}, \tau )  \ne  ( \ve{j}, \tau' )  $. 

Since in the  Kondo phase  $  \ve{\tilde{f}} _{\ve{i}} $    is  the emergent  low lying quasiparticle,  we  expect  the phase 
$\varphi_{\ve{i}}(\tau) $  to  vary slowly  in  time.    This  freezing out  of  the  dynamics  of  the  gauge  field  corresponds  to   
the   Higgs mechanism.   Here   $\varphi$  drops  out  from   the action  and  the   relevant  theory  is  that of  interacting     
$c$  and  $\tilde{f} $  \textit{electrons}   which  is  very  reminiscent  of the physics of the periodic  Anderson model.     This   
formalizes  the   accepted  notion  that  the  Kondo  lattice model   shares  the  very  same physics  as  the  periodic    Anderson model  
in  the local  moment  regime.       If the  ground  state   turns  out to be a Fermi  liquid,  
 then Luttinger  theorem  should apply   and  both  electron  species  should  be included  in the Luttinger  count.  In the  Higgs  phase,  
 $\varphi$,   is  stuck in a    gauge  choice,   say   $\varphi  = 0$.    Hence  there is  no  distinction between  the   
 $\ve{\tilde{f}}_{\ve{i}}(\tau) $  and   $ \ve{f}_{\ve{i}}(\tau)$.    In  other  words,  the   Abrikosov  fermion  looses  its  gauge   
 charge    and    acquires   an   physical  electric  one.   In  this  sense  we  have  for  the Kondo phase
 \begin{equation}
 	 \langle b_{\ve{i}}(\tau) \rangle  \propto  \langle   \ve{f}^{\dagger}_{\ve{i}}(\tau) \ve{c}^{\phantom\dagger}_{\ve{i}}(\tau)  \rangle   \ne 0. 
 \end{equation}   
 Let  us  note  that in any   exact  evaluation  of  the partition function -- as  carried out  in our  Monte Carlo simulations --   
 $   \langle   \ve{f}^{\dagger}_{\ve{i}}(\tau) \ve{c}^{\phantom\dagger}_{\ve{i}}(\tau)  \rangle   $  vanishes  identically.   However,  
 the  measurement of  
   $   \langle   e^{-i \varphi_{\ve{i}(\tau)}} \ve{f}^{\dagger}_{\ve{i}}(\tau) \ve{c}^{\phantom\dagger}_{\ve{i}}(\tau)  \rangle   $   
   is  finite  and   captures  the  hybridization   matrix  element   characteristic  of  large-$N$  mean-field    theories. 

 We  now  argue  that,  at  least in the large-$N$  limit,     $\ve{\tilde{f}}_{\ve{i}}  $  corresponds   to  the composite  fermion  operator.  
Including  amplitude  fluctuations  we  have:
\begin{equation}
	    \ve{\tilde{f}} _{\ve{i}}  \propto   b_{i}(\tau)  \ve{f}_{\ve{i}}(\tau) \propto    
	     \left[ \ve{f}^{\dagger}_{\ve{i}}(\tau) \ve{c}^{\phantom\dagger}_{\ve{i}}(\tau)   \right] 
	     \ve{f}_{\ve{i}}(\tau).
\end{equation}
The  above  is  precisely  the form of the    composite  fermion operator, considered  in this  article,  in the large-$N$ limit~\cite{Danu21}.

  We   are  now in a  position to   define  precisely  the   relevant  phases  of the KLM  that we  encounter in this  article.  
  They  are  summarized in   Table~\ref{tb:phases}.

\begin{table} 
{
\centering
\begin{tabular}{l*{5}{c}r}
\hline
Phases      &~~ $  \langle  \ve{f}^\dagger_{\ve{i}} \ve{ \sigma }  \ve{f}_{i} \rangle$ &~~ &  $ \langle  b_{\ve{i}} (\tau)  \rangle$   \\
\hline\\

SDW         &~~  $\checkmark $       &~~  & $\mathcal{\times}$    \\\\

Kondo       &~~ $   \mathcal{\times}$       &~~  &  $\checkmark $     \\\\
  
Kondo+SDW    &~~   $\checkmark $       &~~  & $ \checkmark $    \\\\
\hline
\hline

\end{tabular}\\
\caption{SDW,    Kondo and  Kondo+SDW  phases  of  the  Kondo lattice model.  $ \checkmark $ ($ \mathcal{\times}$ )   
refers  to a non-vanishing (vanishing) value of the order  parameter. }
\label{tb:phases}
}
\end{table}

\input{QMC_suppl_data.tex}

\input{mean_field.tex}

 \end{document}

%% file: results_new.tex

\begin{figure}
\centering
	\includegraphics[width=0.23\textwidth] {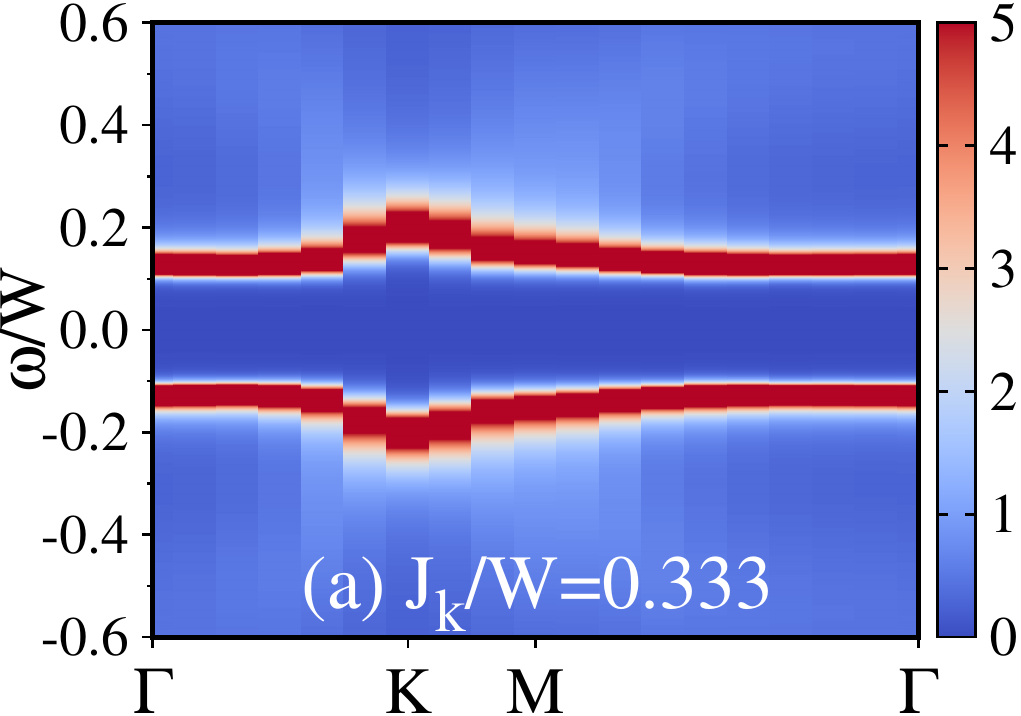}
        \includegraphics[width=0.23\textwidth] {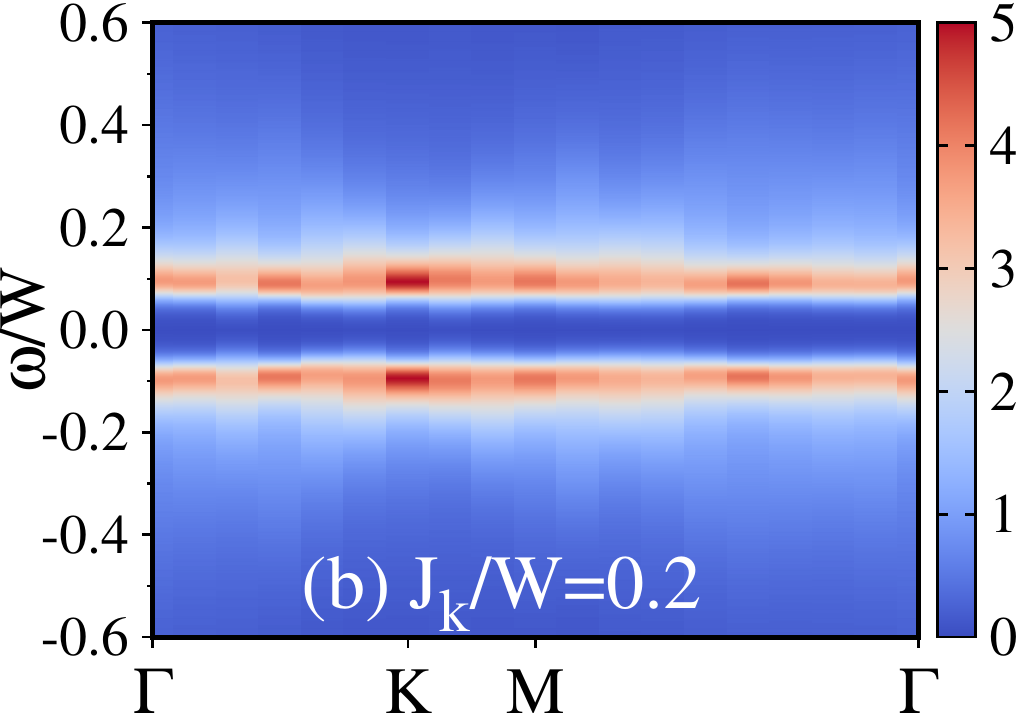}\\
        \includegraphics[width=0.23\textwidth] {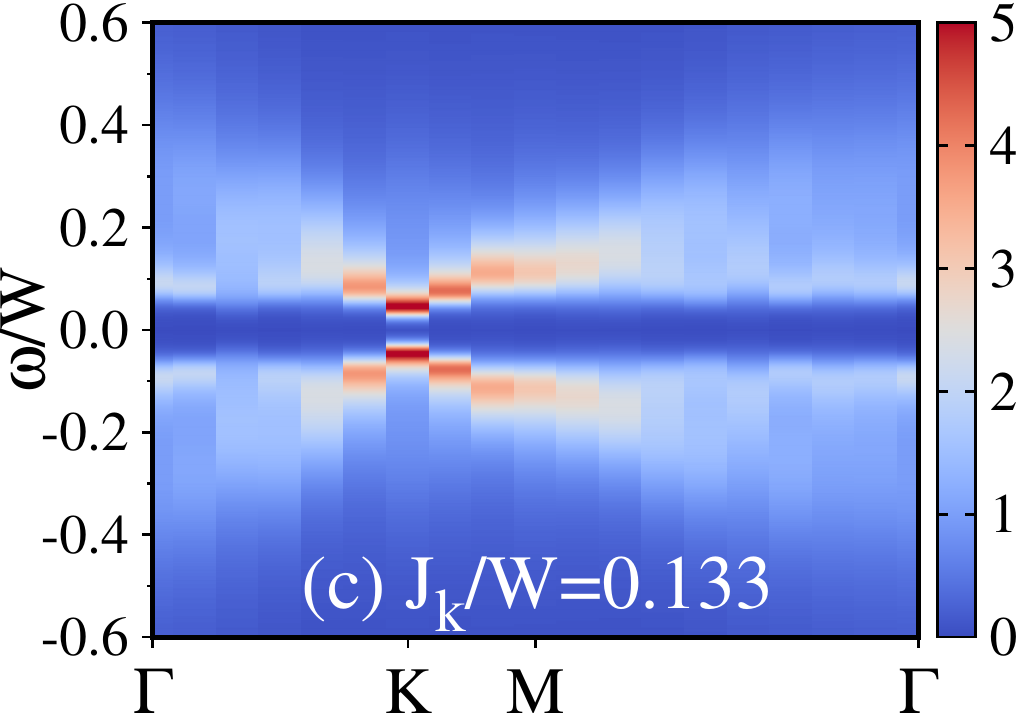}
	\includegraphics[width=0.23\textwidth] {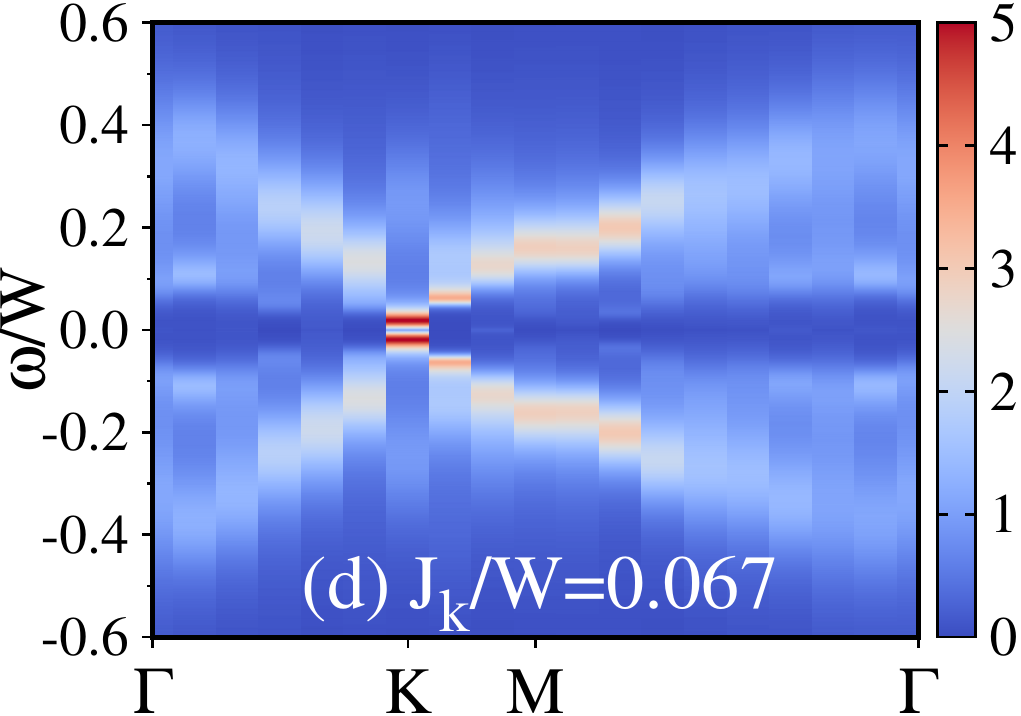}
\caption{Composite fermion spectral function $A_{\psi}(\ve{k},\omega)$ along the $\Gamma$-$K$-$M$-$\Gamma$ path in momentum space with 
        $\Gamma=(0,0)$, $K=(\frac{4\pi}{3},0)$, and $M=(\pi,\frac{\pi}{\sqrt{3}})$ on the $L=18$ honeycomb KLM for representative values 
	of $J_k/W$ corresponding to: (a) Kondo; (b) and (c)  Kondo+SDW, and (d) SDW phases. 
}
\label{fig:AfL18}
\end{figure}

\textit{Results.} 
We first focus on the quantum phase transition between the magnetically ordered and disordered (Kondo) insulators 
and locate the phase boundary by carrying out a finite-size scaling analysis. As detailed in Ref.~\cite{SM}, 
the best data collapse gives the critical value $J_k^c/W=0.2227(3)$ and confirms the expected universality class of the 
three-dimensional classical Heisenberg (O(3)) model.

\begin{figure}[t]
\centering
\includegraphics[width=0.23\textwidth] {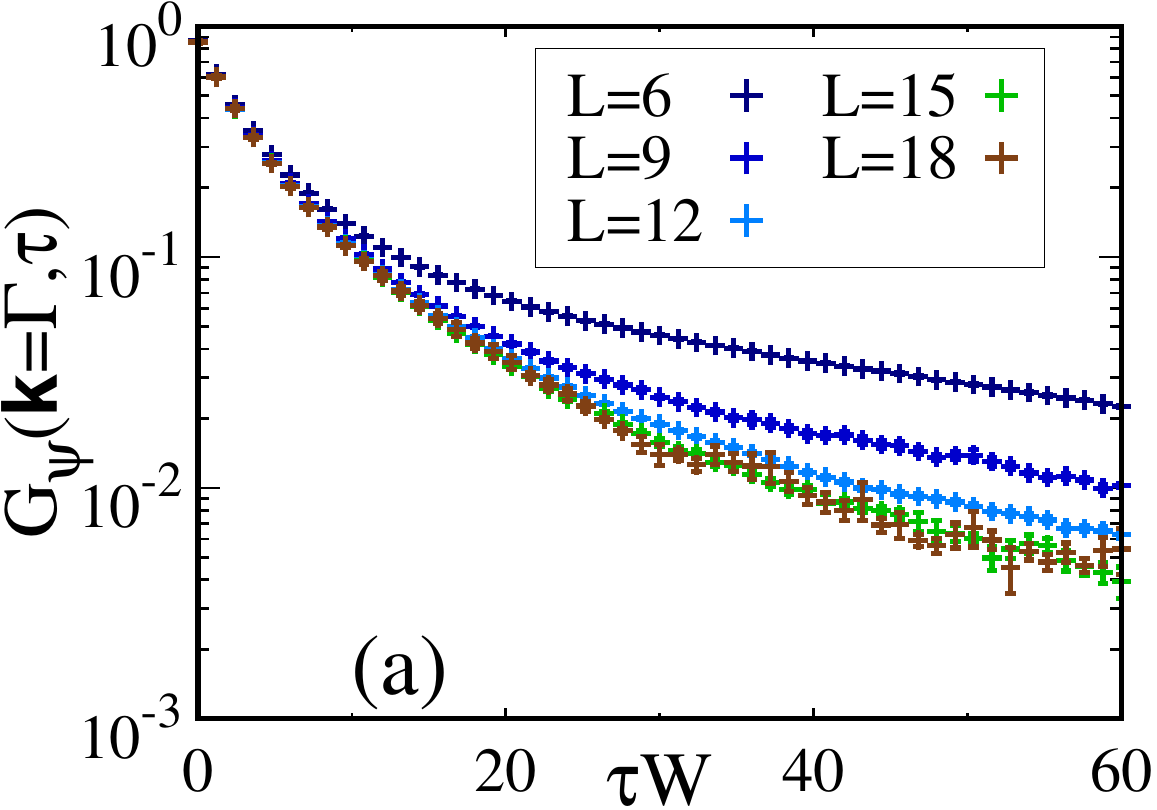}
\includegraphics[width=0.23\textwidth] {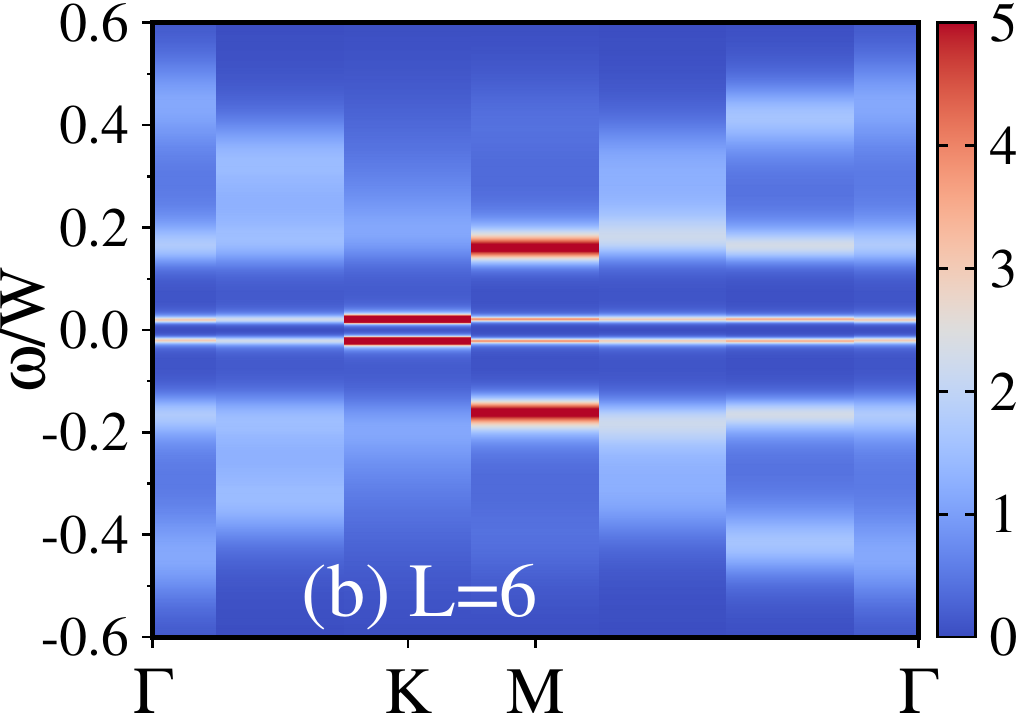}\\
\includegraphics[width=0.23\textwidth] {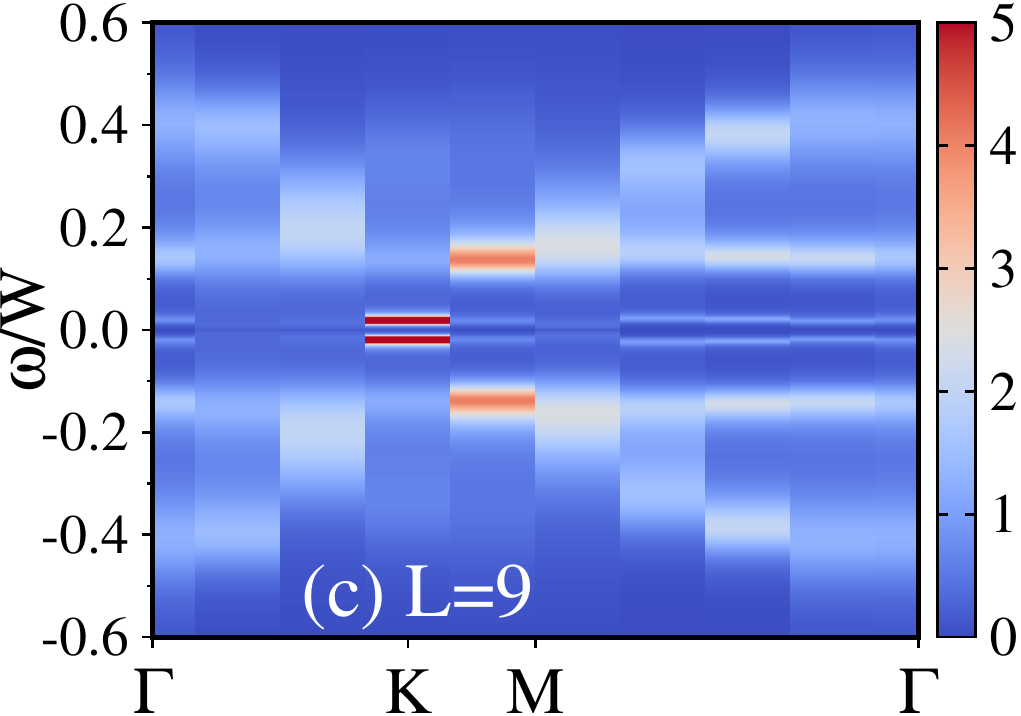}
\includegraphics[width=0.23\textwidth] {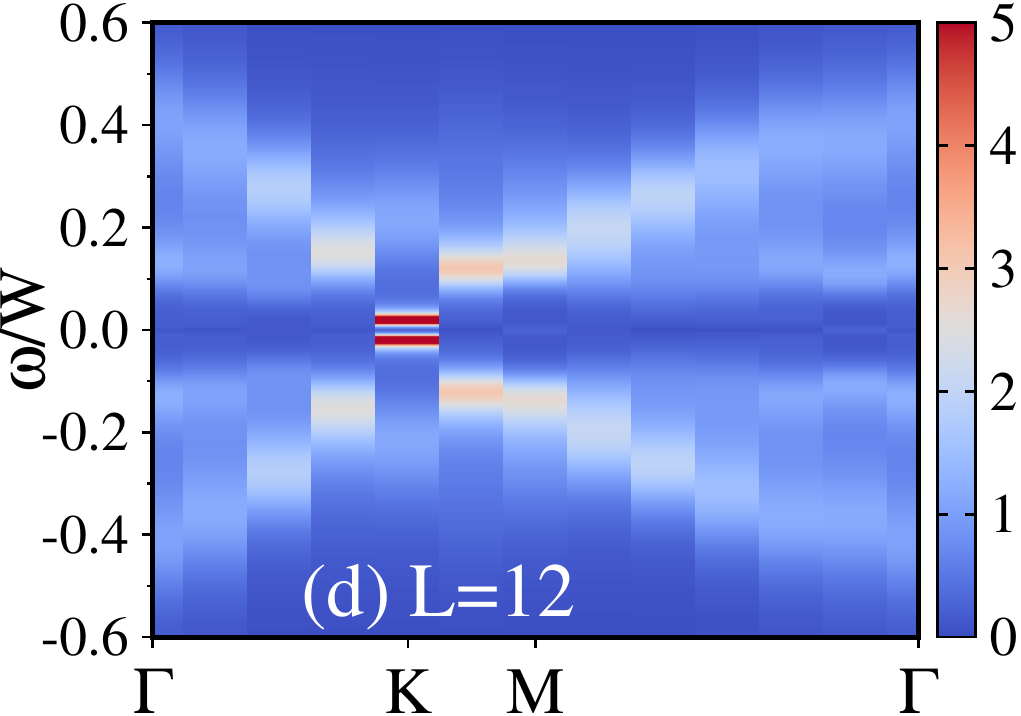}
	\caption{(a) Composite fermion Green's function  $G_{\psi}(\ve{k}=\Gamma,\tau)$ at $J_k/W=0.067$, 
         and (b)-(d)  the corresponding spectral function $A_{\psi}(\ve{k},\omega)$ 
	 on the honeycomb KLM with different sizes $L$. 
}
\label{fig:AfJk04}
\end{figure}

Next, we turn to the evolution of the momentum resolved spectral function of the composite fermion 
$A_{\psi}(\ve{k},\omega) =  - \frac{1}{\pi} \text{Im}    G^{\text ret}_{\psi}(\k,\omega)$ with 
$ G^{\text ret}_{\psi}(\k,\omega)=
	-i \int^\infty_0 dt e^{i \omega t} \sum_{\sigma} 
	\big \langle \big\{ \hat{{\psi}}_{\k,\sigma}(t),\hat{ {\psi}}^\dagger_{\k,\sigma}(0) \big\}  \big\rangle $.
In Fig.~\ref{fig:AfL18}(a) with $J_k/W=0.333$ deep  in the Kondo phase, 
the emergent composite  fermions are clearly manifest as bright weakly dispersive bands throughout the whole irreducible Brillouin zone. 
These bands become less pronounced upon crossing over to the magnetically 
ordered phase, see Figs.~\ref{fig:AfL18}(b) and \ref{fig:AfL18}(c), while some incoherent spectral weight sets in at high energies.   
In contrast, the spectrum in Fig.~\ref{fig:AfL18}(d)  with $J_k/W=0.067$ deep inside the magnetic phase, looks different: 
the composite fermion bands  
have disappeared  indicative of the breakdown of Kondo screening. 
If   Kondo screening  is not present in  the magnetically  ordered  phase,  one can  adopt a large-$S$ approximation. 
In leading order in $S$,  the  spectral  function  $A_{\psi}(\ve{k}, \omega) $  will follow  the conduction electron spectral  
function $A_{c}(\ve{k}, \omega) $, i.e.,   $ A_\psi(\k,\omega) \simeq  S^2 A_c(\k,\omega)$~\cite{Danu21}.    
A comparison of $A_{\psi}(\ve{k}, \omega)$ in Fig.~\ref{fig:AfL18}(d) with the corresponding spectrum $A_{c}(\ve{k}, \omega)$ 
included in Ref.~\cite{SM}, confirms this expectation and allows one to recognize in $A_{\psi}(\ve{k}, \omega)$ 
a pronounced image of the conduction electron band consistent with the large-$S$ limit.

\begin{figure}
\centering
\includegraphics[width=0.23\textwidth] {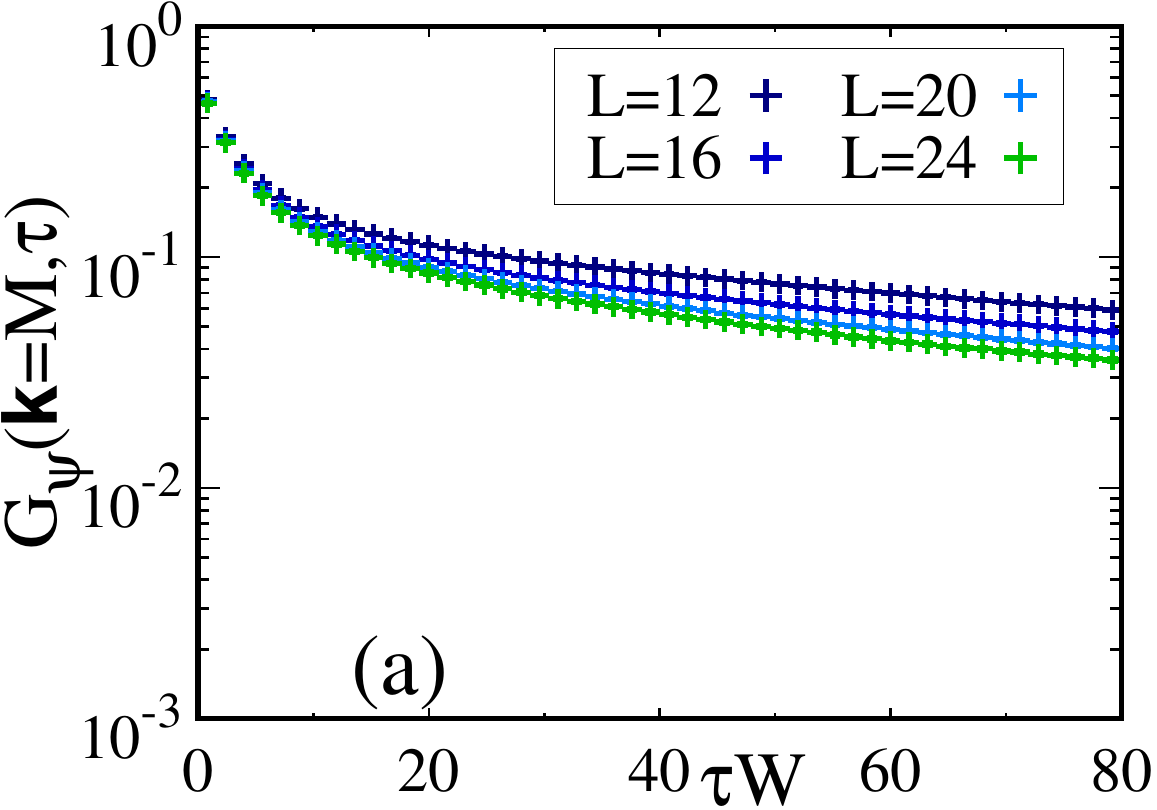}
\includegraphics[width=0.23\textwidth] {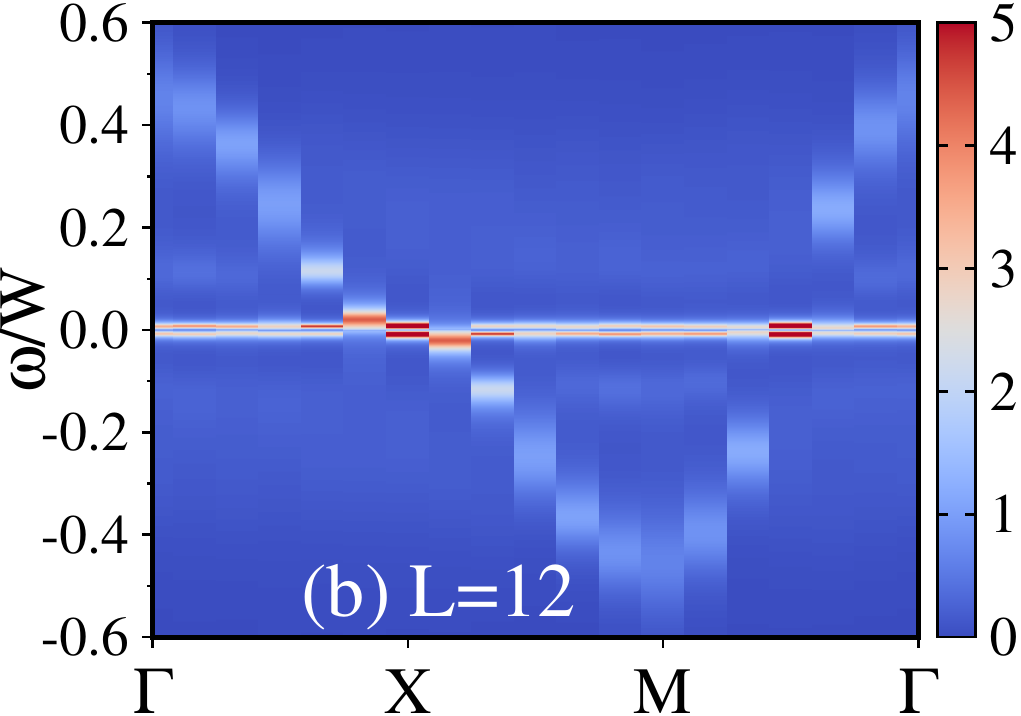}\\
\includegraphics[width=0.23\textwidth] {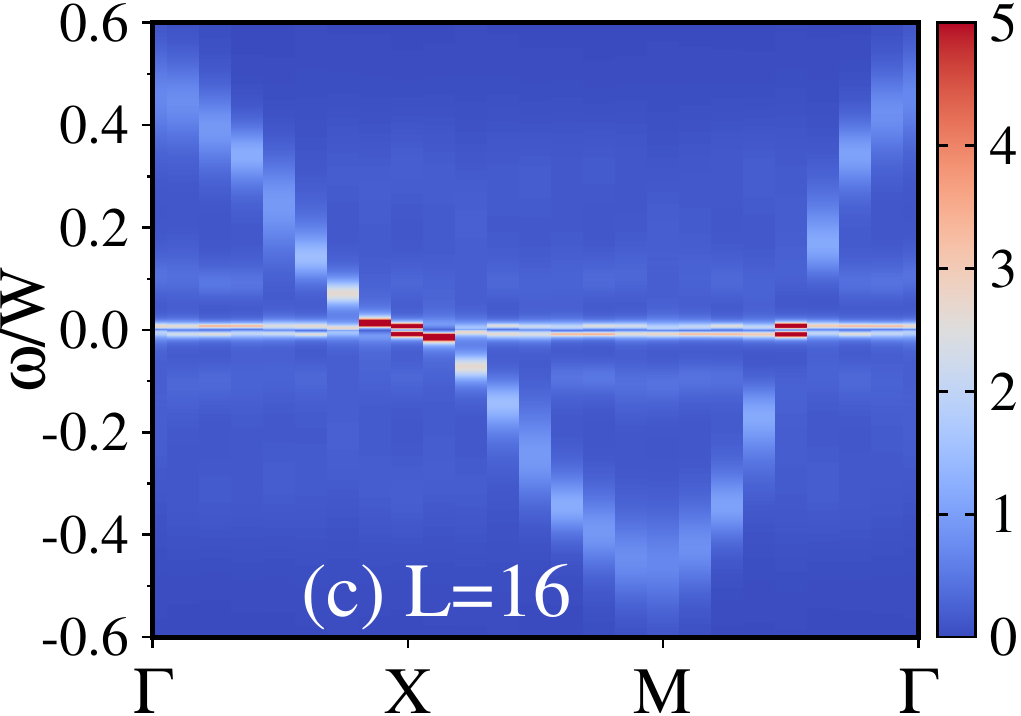}
\includegraphics[width=0.23\textwidth] {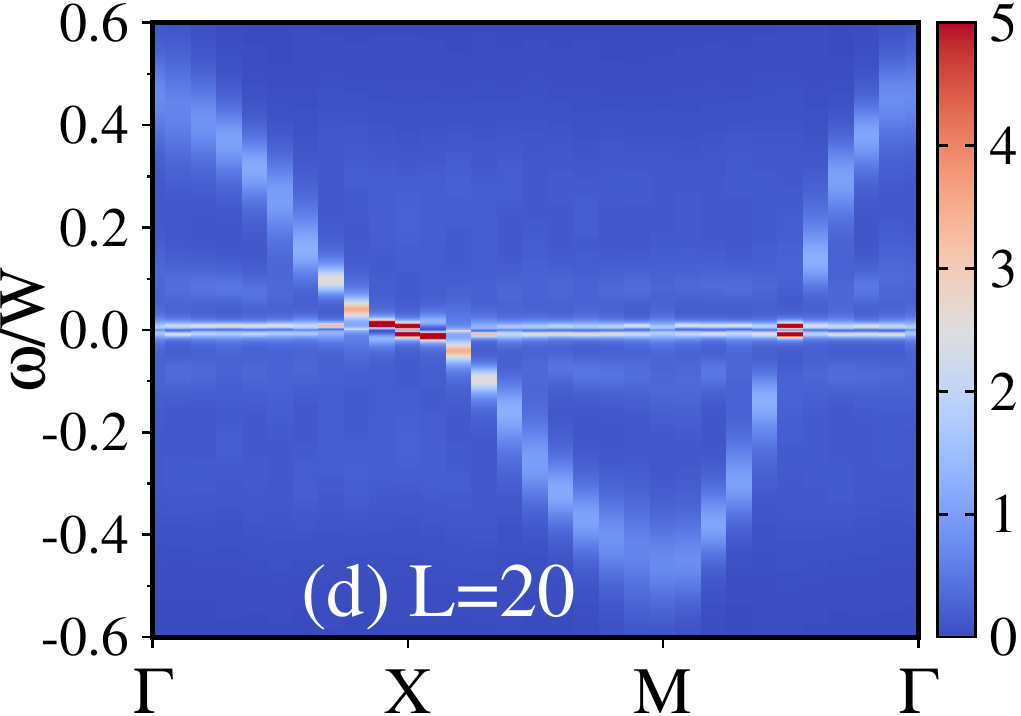}
	\caption{(a) Composite fermion Green's function  $G_{\psi}(\ve{k}=M,\tau)$, where $M=(\pi,\pi)$, at $J_k/W=0.025$,
         and  (b)-(d) the corresponding spectral function $A_{\psi}(\ve{k},\omega)$ along the $\Gamma$-$X$-$M$-$\Gamma$ path, where   
	 $X=(\pi,0)$, on the square KLM with different sizes $L$. 
}
\label{fig:AfJk02}
\end{figure}

In order to get further insight into the observed rearrangement of spectral weight in $A_{\psi}(\ve{k}, \omega)$, 
we plot in Fig.~\ref{fig:AfJk04}(a) raw data of $G_{\psi}(\ve{k},\tau)$ at the $\Gamma$ point 
at our smallest Kondo coupling $J_k/W=0.067$  for different system sizes $L$. 
Generically, the existence of long lived quasiparticles requires that the Green’s function 
displays a free particle behavior at long imaginary times, 
$G_{}(\ve{k},\tau) \stackrel{\tau \to \infty}{\to}  Z^{}_{\ve{k}}e^{-\Delta_{qp}(\ve{k}) \tau }$, 
where $Z^{}_{\ve{k}}$ is the quasiparticle residue of the doped hole at momentum $\ve{k}$ and frequency $\omega=-\Delta_{qp}$. 
As is apparent, the $L=6$ data quickly converge to the exponential decay, which as shown in ~\ref{fig:AfJk04}(b),  
deceptively generates a low energy pole, and consequently a well defined composite fermion band, 
in the corresponding spectral function $A_{\psi}(\ve{k},\omega)$.  
On the other hand, upon increasing system size it becomes more difficult to track the exponential 
form of $G_{\psi}(\ve{k}=\Gamma,\tau)$ whose long time tail systematically flattens.     
As a consequence, while a faint signature  of the composite fermion band can still be spotted in 
$A_{\psi}(\ve{k},\omega)$ for $L=9$, see Fig.~\ref{fig:AfJk04}(c),  
the band has essentially disappeared from the $L=12$ spectrum in Fig.~\ref{fig:AfJk04}(d). 
At the same time, the overall spectrum around the  $\Gamma$ point broadens substantially 
and  may plausibly be thought of as a continuum that stems from decay of the composite quasiparticle.  
Thus, the data are suggestive of the absence of Kondo screening in the thermodynamic limit.

It is striking to compare the results in Fig.~\ref{fig:AfJk04} with those on the square lattice obtained at even smaller 
value of $J_k/W=0.025$, see Fig.~\ref{fig:AfJk02}. Irrespective  of the system size $L$,  the
composite fermion Green's function $G_{\psi}(\ve{k},\tau)$ at the $M=(\pi,\pi)$ point shows the same asymptotic behavior in the long 
time limit which implies the continued existence of the pole in the corresponding spectrum $A_{\psi}(\ve{k},\omega)$,
see Figs.~\ref{fig:AfJk02}(b)-\ref{fig:AfJk02}(d).
As can be seen, $A_{\psi}(\ve{k},\omega)$  shares aspects of both the large-$N$ approach (flat composite fermion bands) and large-$S$ limit, 
i.e., the image of the conduction electron band shifted by the antiferromagnetic wavevector $\ve{Q}=(\pi,\pi)$.  
Taken together, these spectral features imply coexistence of coherent Kondo screening and long range magnetic order.

To substantiate the vanishing of the composite fermion band as a function of $J_k/W$, we extract the quasiparticle residue 
$Z^{\psi}_{\ve{k}}$ at the  $\Gamma$ point by fitting the long time tail of $G_{\psi}(\ve{k}=\Gamma,\tau)$ to the exponential form 
followed by the finite-size scaling analysis~\cite{SM}. For comparison, we have equally analyzed the asymptotic behavior of 
$G_{\psi}(\ve{k},\tau)$ at the $M$ point on the square lattice and constructed the respective phase diagrams compiled in Fig.~\ref{Fig:phase}.   

Since increasing $J_k$ promotes the Kondo effect, it ultimately drives the magnetic order-disorder transition 
that occurs at $J^c_k/W\simeq 0.223$ (honeycomb) and $J^c_k/W\simeq 0.181$ (square)~\cite{Assaad99a,Capponi00,Raczkowski20}. 
Thus, the strong coupling region in Fig.~\ref{Fig:phase} is lattice independent and hosts a Kondo screened phase. 
In contrast, a weak coupling part of the phase diagram turns out to be non-generic:  
While pinning down the precise scaling  of $Z^{\psi}_{\ve{k}}$ at the $\Gamma$ point on the honeycomb lattice is a challenge, 
our data show that it is a monotonically decreasing function of $J_k/W$ and vanishes slightly below $J_k/W=0.1$. 
The vanishing quasiparticle residue indicates that composite quasiparticles lose their integrity. 
We interprete this as the destruction of Kondo screening. 
This is in stark contrast to the square lattice where composite fermions are found  down to our smallest value $J_k/W=0.025$ 
as signaled by a finite quasiparticle residue $Z^{\psi}_{\ve{k}}$ at the $M$ point.      

\begin{figure}
\centering
\includegraphics[width=0.48\textwidth] {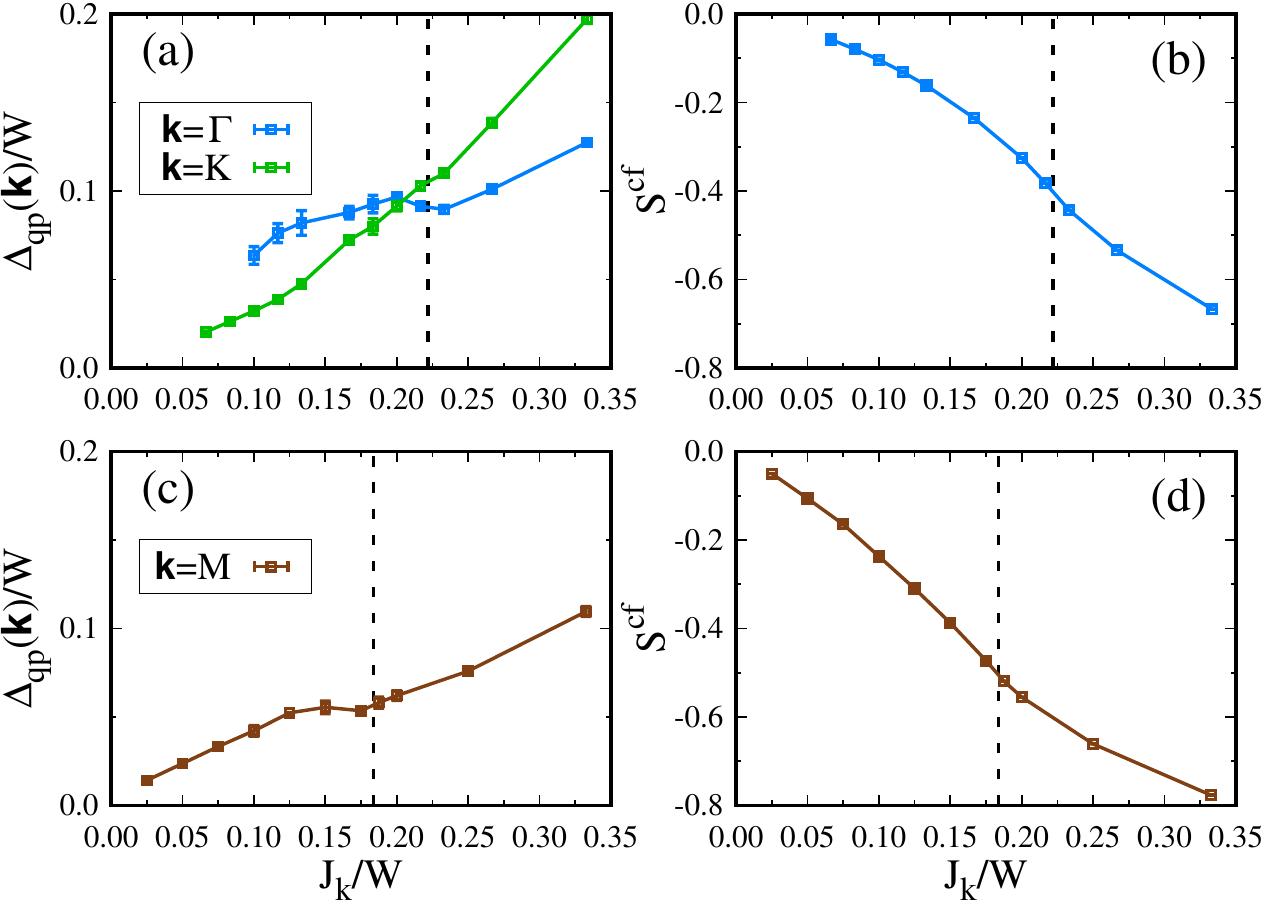}
	\caption{(a) Single particle gap $\Delta_{qp}(\ve{k})$ at the $\Gamma$ and Dirac K points  and (b) 
	the local spin-spin correlation function $S^{cf}=\tfrac{2}{3N}\sum_{\ve{i}}\langle 
        \ve{\hat{c}}^{\dagger}_{\ve{i}} \ve{\sigma} \ve{\hat{c}}^{\phantom\dagger}_{\ve{i}} \cdot \ve{\hat{S}}^{}_{\ve{i}} \rangle$
	as a function of $J_k/W$ on the honeycomb lattice. For comparison,  we show in (c)  $\Delta_{qp}(\ve{k})$ at the $M$ point 
	and (d) $S^{cf}$ on the square lattice.
	Dashed lines denote the respective magnetic order-disorder transitions.  
	All quantities are representative of the thermodynamic limit~\cite{SM}.   
	}
\label{fig:Delta_k}
\end{figure}

We also track the location and the size of the quasiparticle gap.   
Given that at large $J_k/W$ the quasiparticle gap is located at the $\Gamma$ point while the noninteracting model features gapless 
Dirac excitations at the $K$ point, one shall resolve a change in the position of the minimal gap as a function of $J_k/W$. 
The data in Fig.~\ref{fig:Delta_k}(a) extracted from the long time behavior of $G_{\psi}(\ve{k},\tau)$ 
at the both $\ve{k}$ points confirm this expectation. As is apparent, the change takes place on the magnetically ordered side of $J_k^c$ but 
far away from  Kondo breakdown. Further, the comparison of Figs.~\ref{fig:Delta_k}(a) and \ref{fig:Delta_k}(c), the latter 
showing the evolution of the quasiparticle gap at the $M$ point on the square lattice, reveals two common features: 
(i) the development of the cusp preceding the magnetic order-disorder transition, and 
(ii) a linear in $J_k/W$  scaling of the gap in the weak coupling limit. It is a direct consequence of the Fermi surface nesting-driven 
magnetic order and can be captured within a mean-field SDW framework~\cite{EPJB.86.195,PhysRevB.62.76}.

Finally, as shown in Fig.~\ref{fig:Delta_k}(b)  we do not resolve any signs of the breakdown of Kondo screening in the 
local spin-spin correlation function 
$S^{cf}=\tfrac{2}{3N}\sum_{\ve{i}}\langle
        \ve{\hat{c}}^{\dagger}_{\ve{i}} \ve{\sigma} \ve{\hat{c}}^{\phantom\dagger}_{\ve{i}} \cdot \ve{\hat{S}}^{}_{\ve{i}} \rangle$ 
which remains finite down to our lowest value of $J_k/W$, just like that measured on the square lattice, see Fig.~\ref{fig:Delta_k}(d). 
This seemingly counterintuitive result becomes clear by noting that  $S^{cf}$ measures the amplitude  of the  boson  field, $|b|^2$.  
Hence,  Fig.~\ref{fig:Delta_k}(b)  implies  that the modulus of the boson field  remains  constant  for  all values of  
the Kondo coupling and that  Kondo  breakdown  occurs  due  to phase  fluctuations. 
The latter  explains the failure of the mean-field approaches to provide consistent results for both lattices~\cite{SM}.

%% file: QMC_suppl_data.tex
\section{Supplemental QMC results}

\subsection{QMC setup}

The approach  relies  on the U(1)  gauge  formulation of the KLM  described  above.
The  integration over  the   Grassmann  variables  yields  the  fermion determinant.  
For  the  particle-hole   symmetric conduction band,  one  will readily  show,  that  it is positive semi-definite~\cite{Wu04,PhysRevLett.120.107201}.
To  formulate  the  algorithm,  we  discretize  the  imaginary time   and  choose  $\Delta  \tau t  =  0.2$ ($\Delta  \tau t  =  0.1$)
on the honeycomb (square) lattice   and use  the Gauss-Hermite   quadrature  to   discretize  the  fields.

We  have used  a  projective   version of  the QMC algorithm  based on the imaginary time evolution of a trial wave function 
$| \Psi_\text{T}\rangle$, with $ \langle\Psi_\text{T}  |\Psi_0 \rangle  \neq 0 $,  to the ground state $|\Psi_0 \rangle$:         
\begin{equation}
  \frac{ \langle  \Psi_0 | \hat{O} |  \Psi_0 \rangle  }{ \langle  \Psi_0  |  \Psi_0 \rangle  }  =  
  \lim_{\Theta \rightarrow \infty} 
  \frac{ \langle  \Psi_\text{T} | e^{-\Theta \hat{\mathcal{H}}  } \hat{O} e^{-\Theta \hat{\mathcal{H}}   } |  \Psi_\text{T}\rangle
  }
  { \langle  \Psi_\text{T}  | e^{-2\Theta \hat{\mathcal{H}} }  |  \Psi_\text{T} \rangle  }.
\end{equation}
Since the energy scale of the RKKY interaction scales as $J_k^2$, convergence to the magnetically ordered ground state in the weak coupling  
requires adequately increased projection parameters, i.e., $\Theta t=40 $ at $J_k/t=0.8$ and  $\Theta t=160 $ at $J_k/t=0.4$ 
on the honeycomb lattice. On the other hand, on the square lattice $\Theta t=80$ was found to be already sufficient down to $J_k/t=0.2$.    

For  the analytical continuation,  we have made  use of the  stochastic Maximum Entropy  method~\cite{Beach04a}  implemented in the ALF-library.

\subsection{Magnetic order-disorder transition}

\begin{figure}
\centering
\includegraphics[width=0.48\textwidth] {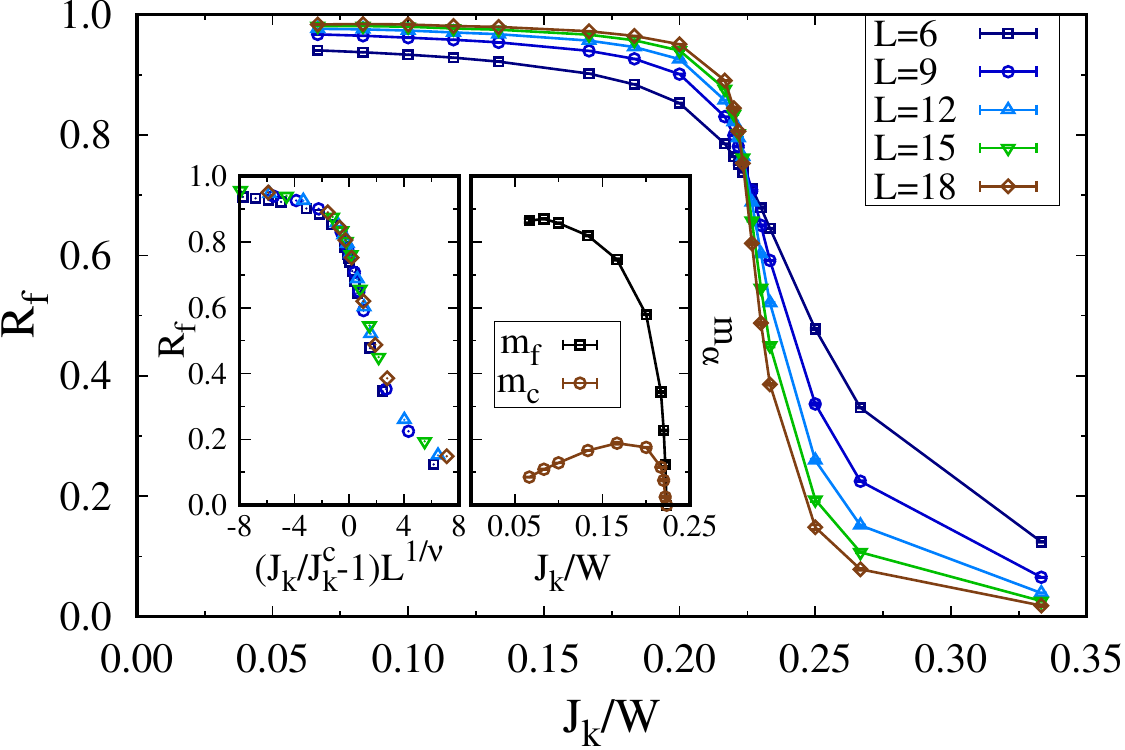}
	\caption{Correlation ratio $R_f$  defined in Eq. (\ref{eq:R}) as a function of $J_k/W$.  
	Left inset shows the scaling collapse of $R_f$ for $L\geq 12$ 
	assuming the critical exponent $1/\nu=1.40511(6)$ of the 3D classical Heisenberg (O(3)) model 
	\cite{PhysRevD.104.105013}. 
	Right inset shows the staggered magnetic moment $m_{\alpha=\{c,f\}}$ in the thermodynamic limit.  
}
\label{fig:R}
\end{figure}

In order to determine the precise location of the magnetic order-disorder transition, we calculate the spin structure factor 
for the $f$ spins
\begin{equation}
        S^{f}(\ve{k})=  
        \frac{4}{L^2} \sum_{\delta=A,B} \sum_{\ve{r}} e^{i\ve{k}\cdot \ve{r}} 
        \langle \hat{\ve{S}}^{}_{\delta}(\ve{r})\cdot \hat{\ve{S}}^{}_{\delta} \rangle, 
\label{eq:Sf}
\end{equation}
from which we construct the renormalization group invariant correlation ratio
\begin{equation}
        R_{f} =  1 - \frac{S^{f}(\ve{Q} + \delta\ve{k}) }{S^{f}(\ve{Q}) },
\label{eq:R}
\end{equation}
where $\ve{Q}=(0,0)$ is the ordering wavevector and $\delta\ve{k}$ is the smallest wavevector on the $L\times L$ honeycomb lattice.
As can be seen in Fig.~\ref{fig:R}, $R_f$ scales to unity (zero) for ordered (disordered) 
states and shows a crossing point as a function of system size at the critical point $J_k^c$. 
Given that the charge degrees of freedom are gapped across the transition, we expect that it belongs to the universality class of 
the  3D classical Heisenberg (O(3)) model. Indeed, assuming the correlation length exponent $1/\nu=1.40511(6)$ 
\cite{PhysRevD.104.105013} of the latter and using the scaling assumption
\begin{equation}
R_f=f[(J_k/J_k^c-1)L^{1/\nu}], 
\end{equation}
we obtain for $L\geq 12$ a good quality data collapse of $R_f$ shown in the left inset of Fig.~\ref{fig:R}. 
It allows us to estimate $J_k^c/W=0.2227(3)$.

\begin{figure*}[h!]
\begin{center}
        \includegraphics[width=0.32\textwidth]{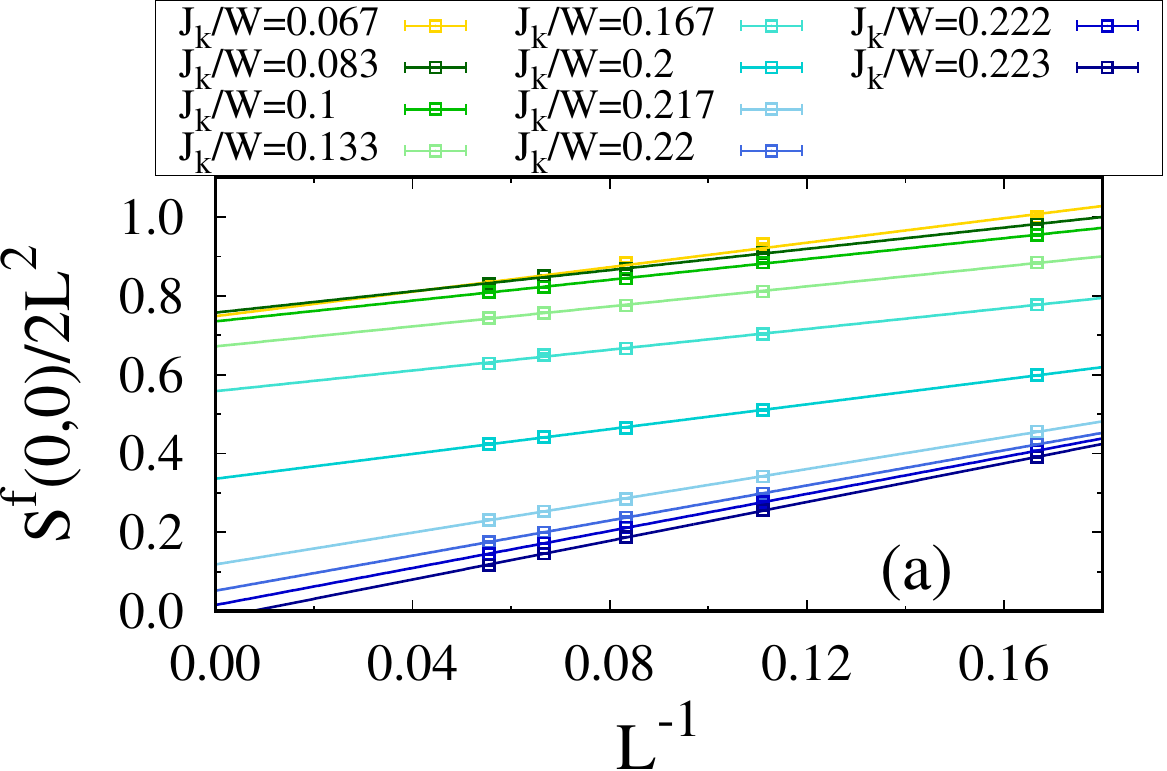}
        \includegraphics[width=0.32\textwidth]{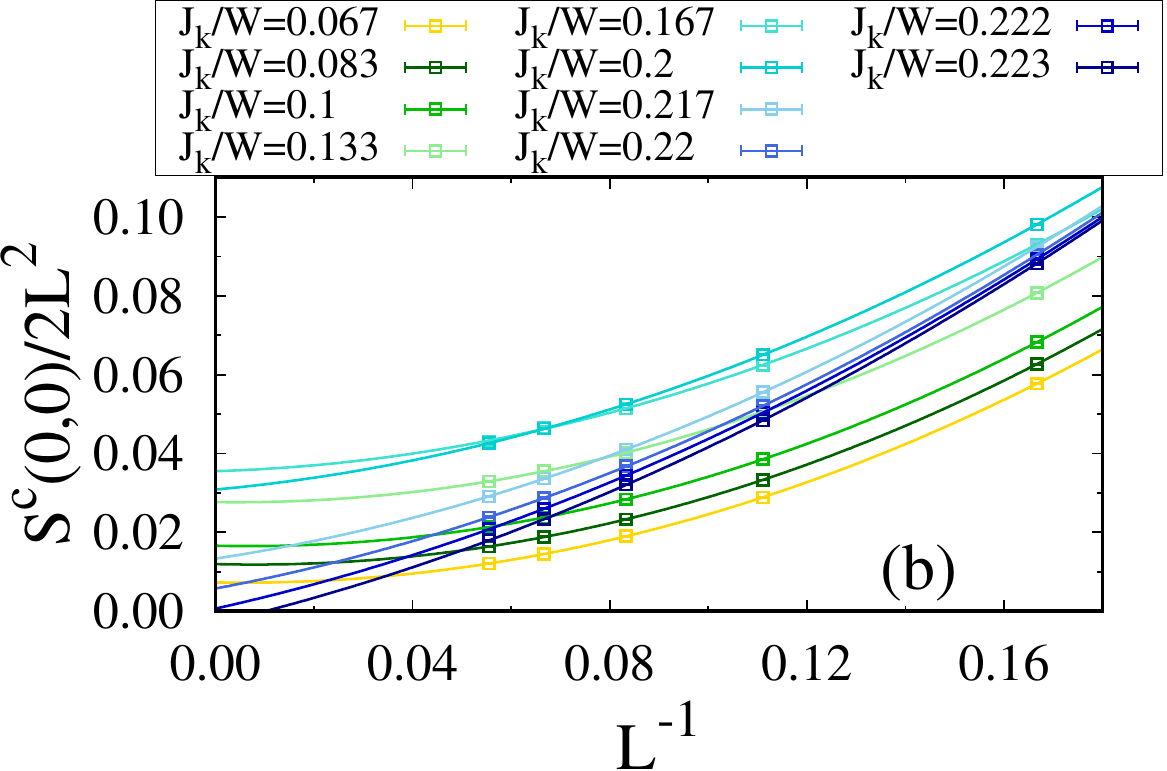}
\end{center}
\caption{Finite-size extrapolation of the antiferromagnetic spin structure factor at $\ve{Q}=(0,0)$ on the honeycomb KLM  
	for the (a) $f$- and (b)  $c$-electrons on approaching the magnetic order-disorder transition point; 
	solid lines are linear and second-order polynomial fits to the QMC data. 
}
\label{S:SQ}
\end{figure*}

As shown in Fig.~\ref{S:SQ},  we have equally performed finite-size scaling of both $S^{f}(\ve{Q})$ 
and the spin structure factor for the conduction electron spins
 \begin{equation}
	S^{c}(\ve{Q})=
        \frac{1}{L^2} \sum_{\delta=A,B} \sum_{\ve{r}} e^{i\ve{Q}\cdot \ve{r}}
	\langle \ve{\hat{c}}^{\dagger}_{\delta}(\ve{r}) \ve{\sigma} \ve{\hat{c}}^{\phantom\dagger}_{\delta}(\ve{r}) \cdot
	 \ve{\hat{c}}^{\dagger}_{\delta} \ve{\sigma} \ve{\hat{c}}^{\phantom\dagger}_{\delta} \rangle.
\end{equation}
We have used linear [Fig.~\ref{S:SQ}(a)] and second-order [Fig.~\ref{S:SQ}(b)] polynomial forms in $1/L$. 
The resultant orbital $\alpha=\{c,f\}$  resolved staggered magnetic moments
\begin{equation}
        m_{\alpha}  =\sqrt{  \lim_{L\to \infty} \frac{S^{\alpha} (\ve{Q})} {2L^2} }
\end{equation}
both scale continuously to zero at $J_k^c/W=0.223(1)$ (see the right inset of Fig.~\ref{fig:R})  which matches perfectly the previously 
extracted critical point $J_k^c/W=0.2227(3)$.  We  note  that the  good  agreement  between the  extrapolation with  an  analytical  form 
in  $1/L$  and  the data  collapse  based on the  correlation  ratio $R_f$ can be ascribed  to the  very small  anomalous   dimension of
3D O(3)   criticality.

\subsection{Conduction electron spectral function}

\begin{figure}[h]
\centering
        \includegraphics[width=0.23\textwidth] {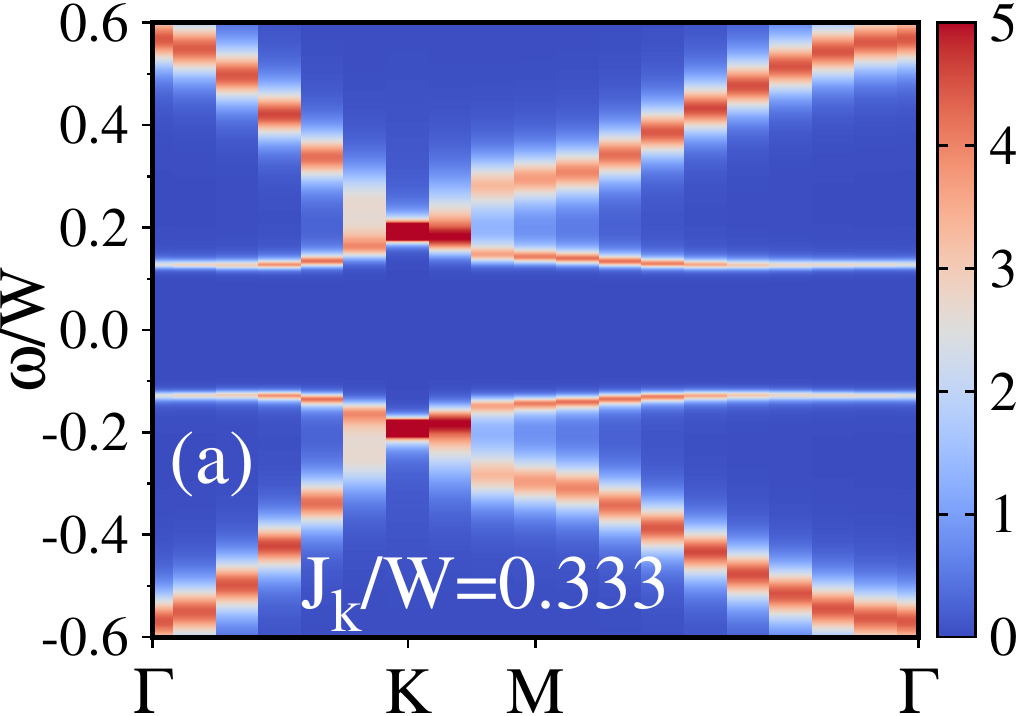}
        \includegraphics[width=0.23\textwidth] {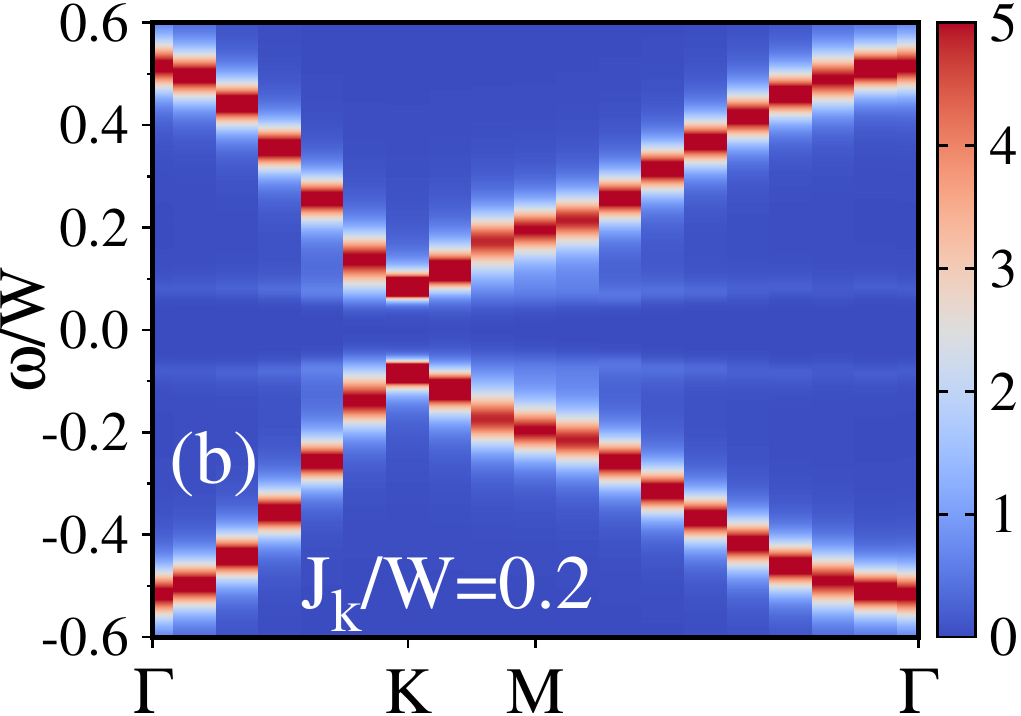}\\
        \includegraphics[width=0.23\textwidth] {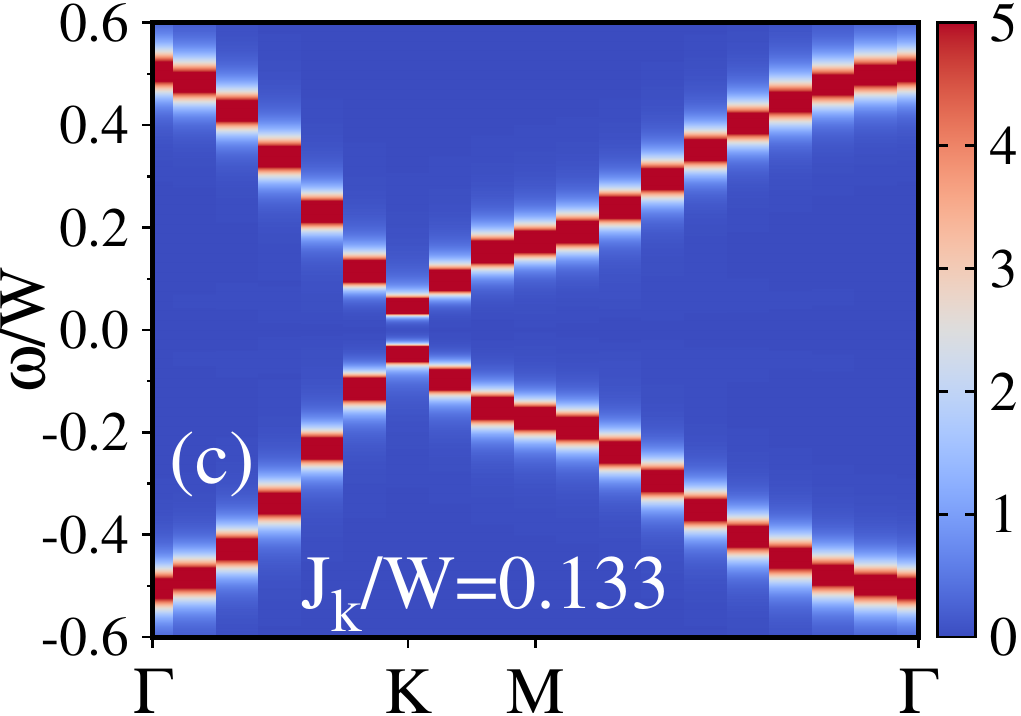}
        \includegraphics[width=0.23\textwidth] {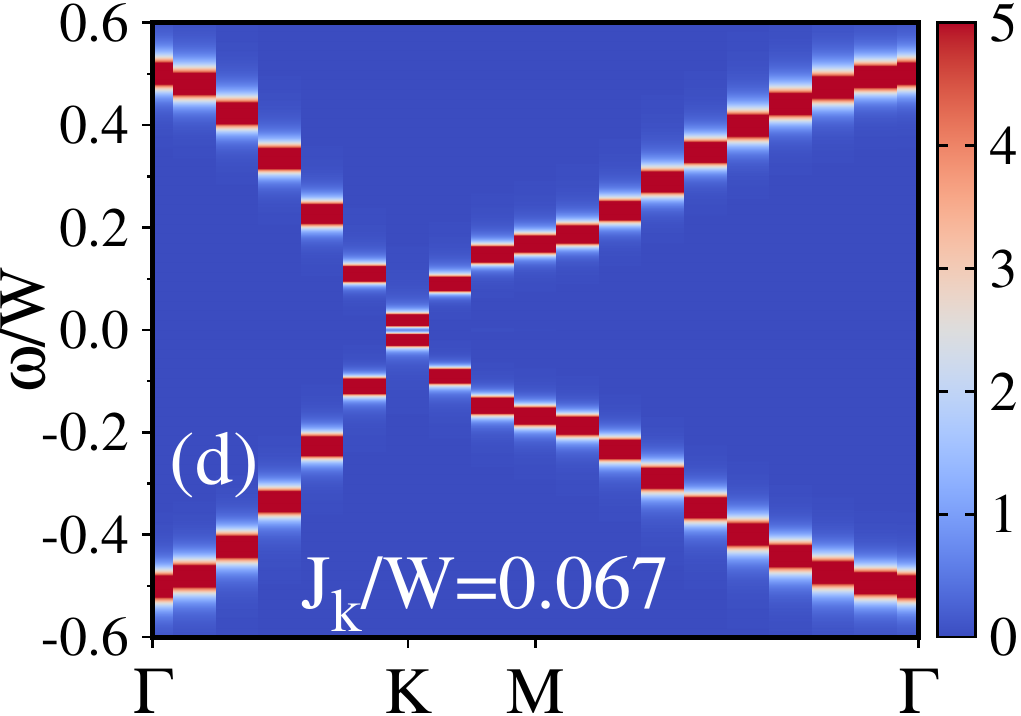}
	\caption{Same as in Fig.~\ref{fig:AfL18} in the main text but for the conduction electrons. 
}
\label{S:AcL18}
\end{figure}

Figure~\ref{S:AcL18} plots the conduction electron  spectral function,
$A_c(\k,\omega) =-\frac{1}{\pi} ~\text{Im}  ~ G^{\text ret}_c(\k,\omega)$, where
\begin{equation}
 G^{\text ret}_c(\k,\omega)=-i \int^{\infty}_0 d t e^{i \omega t} 
	 \sum_{\sigma} \big \langle \big\{ {\hat c}_{\k,\sigma}(t), {\hat c}^\dagger_{\k,\sigma}(0) \big\} \big\rangle
\end{equation}
from the stochastic analytical continuation of the QMC data generated on the $L=18$ honeycomb KLM.

In the region of the phase diagram with active Kondo screening, the composite fermion  and  conduction electron operators 
share the same quantum numbers. Thus their single particle spectral functions shall have identical supports both revealing 
the low energy composite fermion bands. However, the corresponding quasiparticle poles of the conduction electron Green's function 
carry much less spectral weight such that $A_c(\k,\omega)$   exhibits relatively faint  bands, 
see Figs.~\ref{S:AcL18}(a) and \ref{S:AcL18}(b). Moreover, since the quasiparticle  residue  in the small  $J_k/W$ limit  
tracks the  Kondo   scale   $ Z_{\ve{k}}  \simeq e^{-W/J_k} $, resolving composite quasiparticles 
in Fig.~\ref{S:AcL18}(c) requires a logarithmic scale as the spectral weight is nearly fully exhausted by two bands separated by a small 
gap at the Dirac point $K$ but otherwise closely reminiscent of  the tight binding band structure of the honeycomb lattice.

\subsection{Quasiparticle residue $Z^{\psi}_{\ve{k}}$ and single particle gap $\Delta_{qp}(\ve{k})$}

\begin{figure*}[t!]
\begin{center}
	\includegraphics[width=0.32\textwidth]{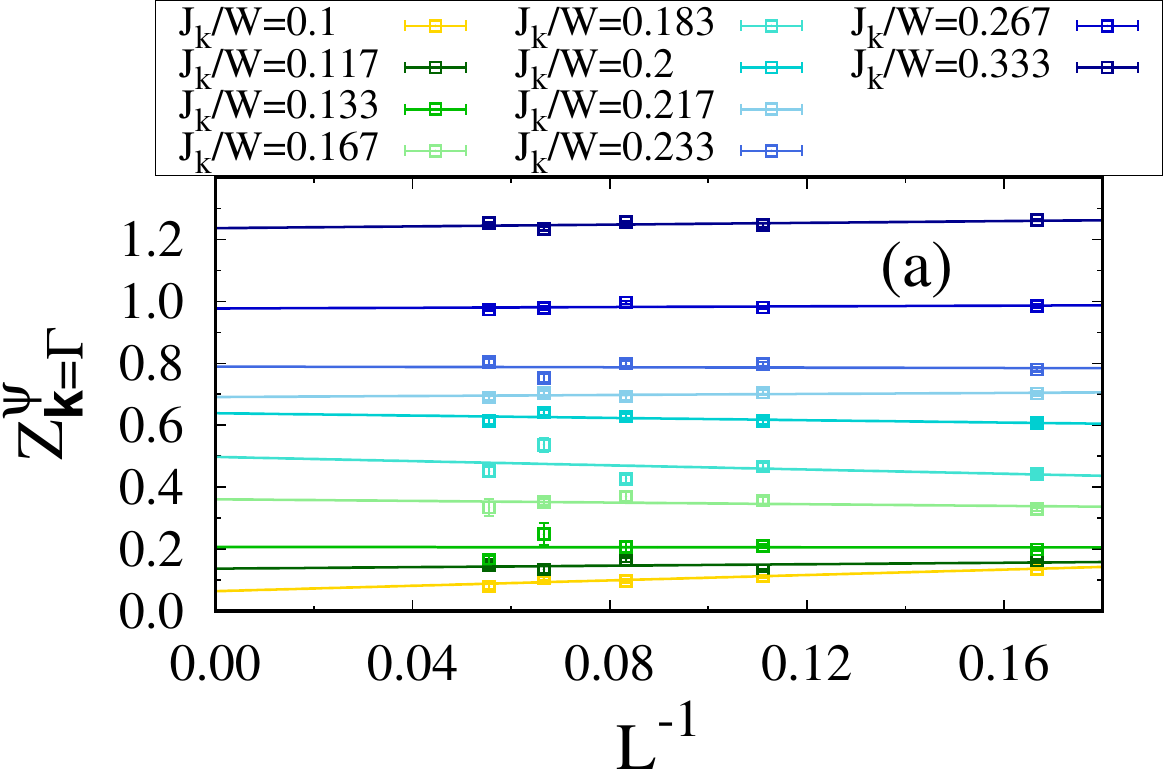} 
	\includegraphics[width=0.32\textwidth]{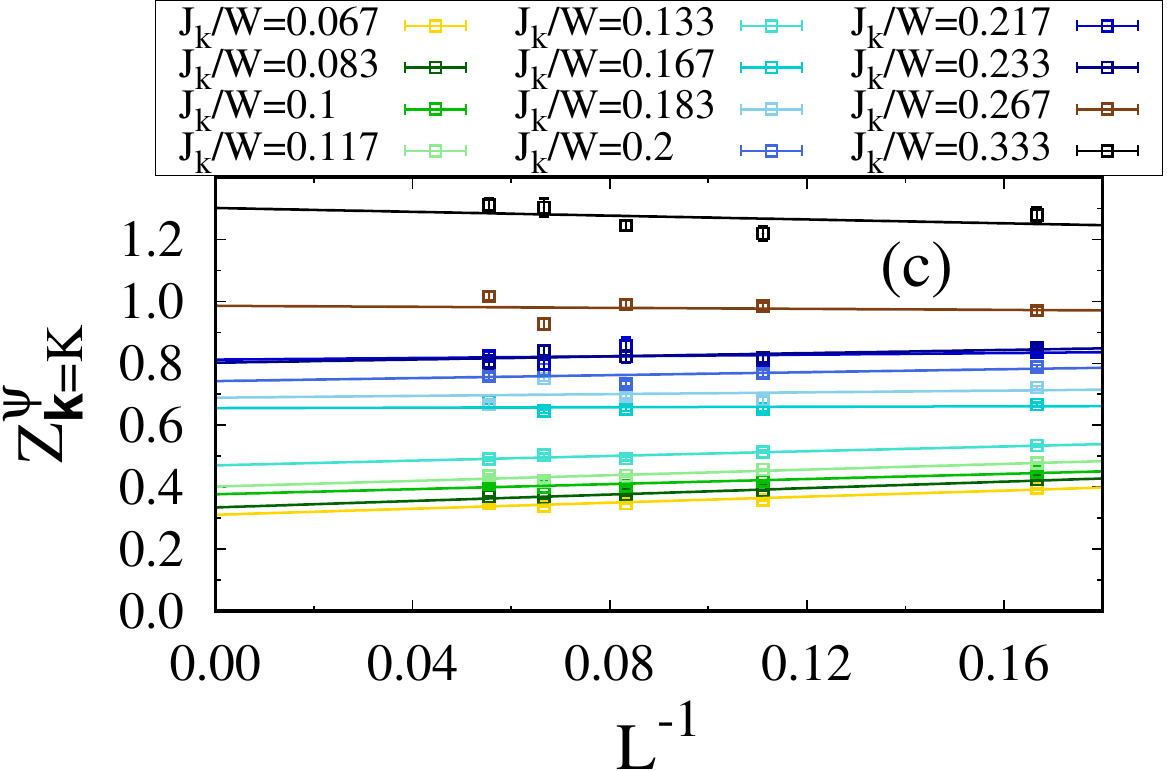} 
	\includegraphics[width=0.32\textwidth]{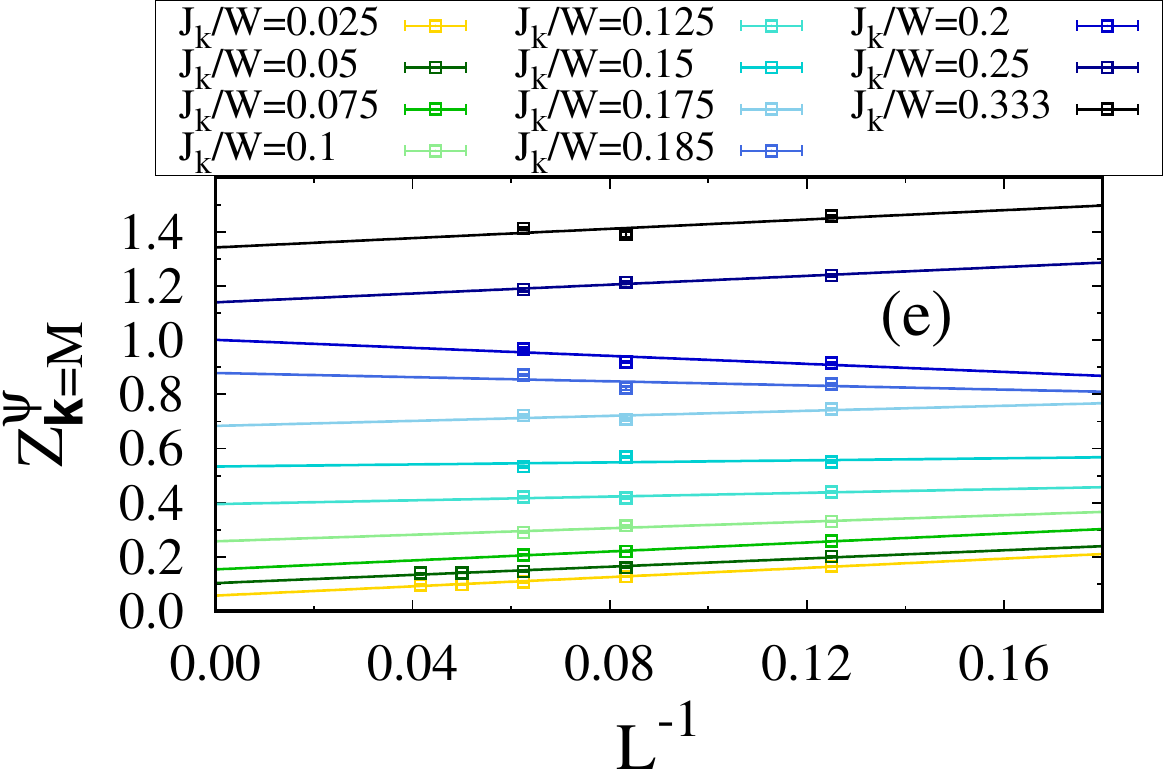}\\
	\includegraphics[width=0.32\textwidth]{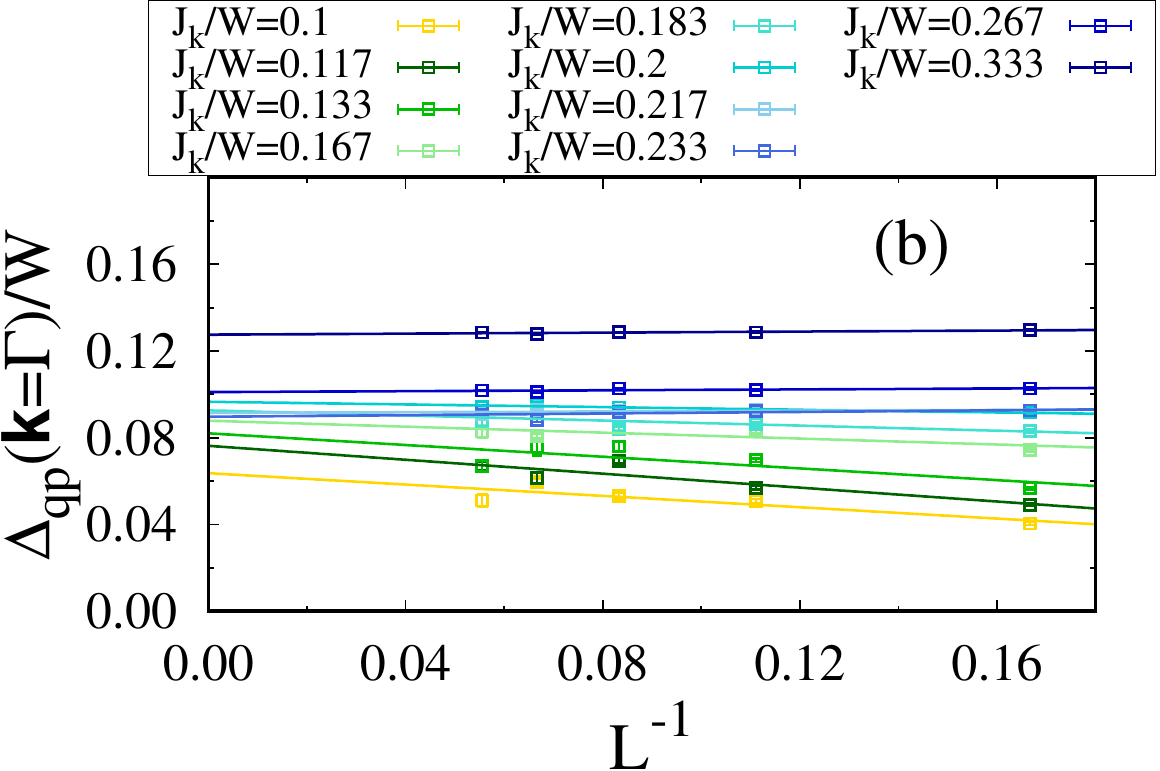}
        \includegraphics[width=0.32\textwidth]{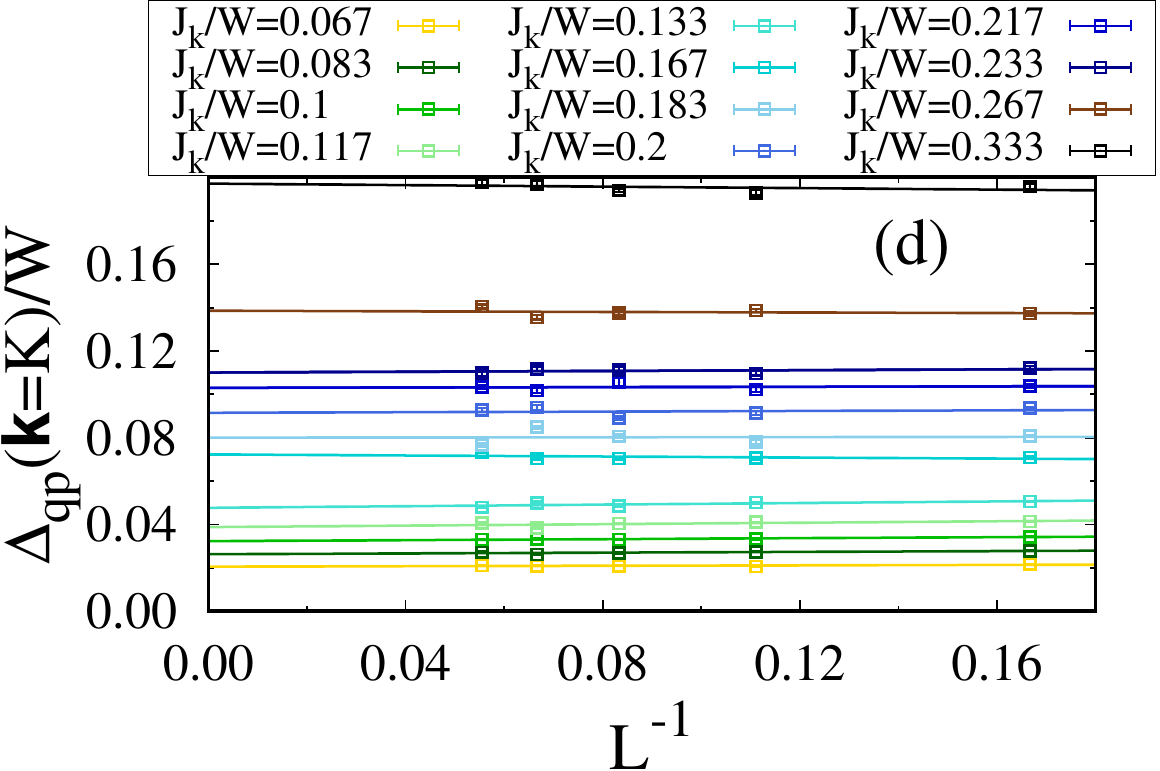}	 
	\includegraphics[width=0.32\textwidth]{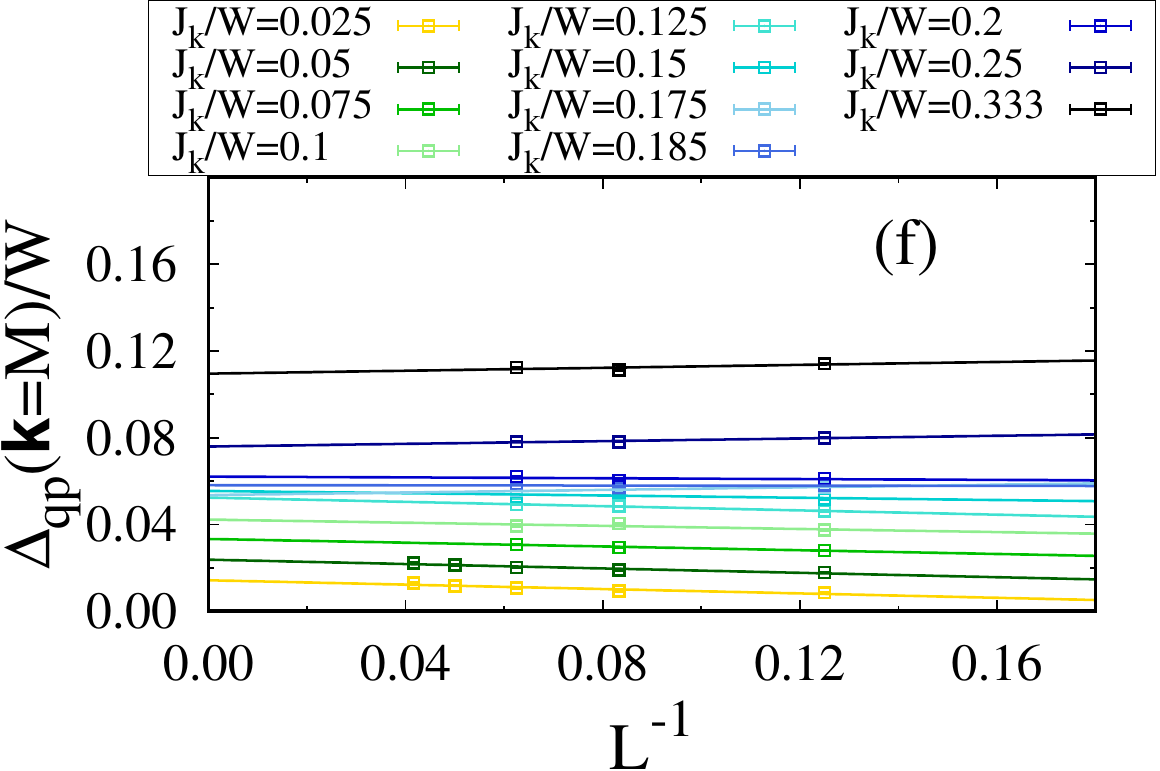}
\end{center}
\caption{Finite-size extrapolation of the quasiparticle residue $Z^{\psi}_{\ve{k}}$ (top) and the corresponding single particle 
        gap $\Delta_{qp}(\ve{k})$ (bottom) extracted from the imaginary-time composite fermion Green's function  $G_{\psi}(\ve{k},\tau)$ 
	at the (a,b) $\Gamma=(0,0)$ and (c,d)  Dirac $K=(\frac{4\pi}{3},0)$ points on the honeycomb KLM and at the (e,f) $M=(\pi,\pi)$ point 
	on the square KLM. Solid lines are linear in $1/L$ fits to the QMC data.
}
\label{S:Zk}
\end{figure*}

The behavior of the imaginary time composite fermion Green's function  $G_{\psi}(\ve{k},\tau)$  at large times, 
$G_{}(\ve{k},\tau) \stackrel{\tau \to \infty}{\to}  Z^{}_{\ve{k}}e^{-\Delta_{qp}(\ve{k}) \tau }$, allows one to extract 
the quasiparticle residue $Z^{\psi}_{\ve{k}}$  and the corresponding single particle gap $\Delta_{qp}(\ve{k})$ without 
the need of analytical continuation.
Figure~\ref{S:Zk} shows the finite-size scaling analysis of the resultant QMC data which led us to the $J_k/W$-dependence of  
the quasiparticle residue $Z^{\psi}_{\ve{k}}$ and the corresponding  single particle gap $\Delta_{qp}(\ve{k})$ 
at the $\Gamma$ and Dirac $K$ points (honeycomb lattice) as well as at the $M$ point (square lattice) presented in the main text.

\begin{figure}
\centering
        \includegraphics[width=0.48\textwidth] {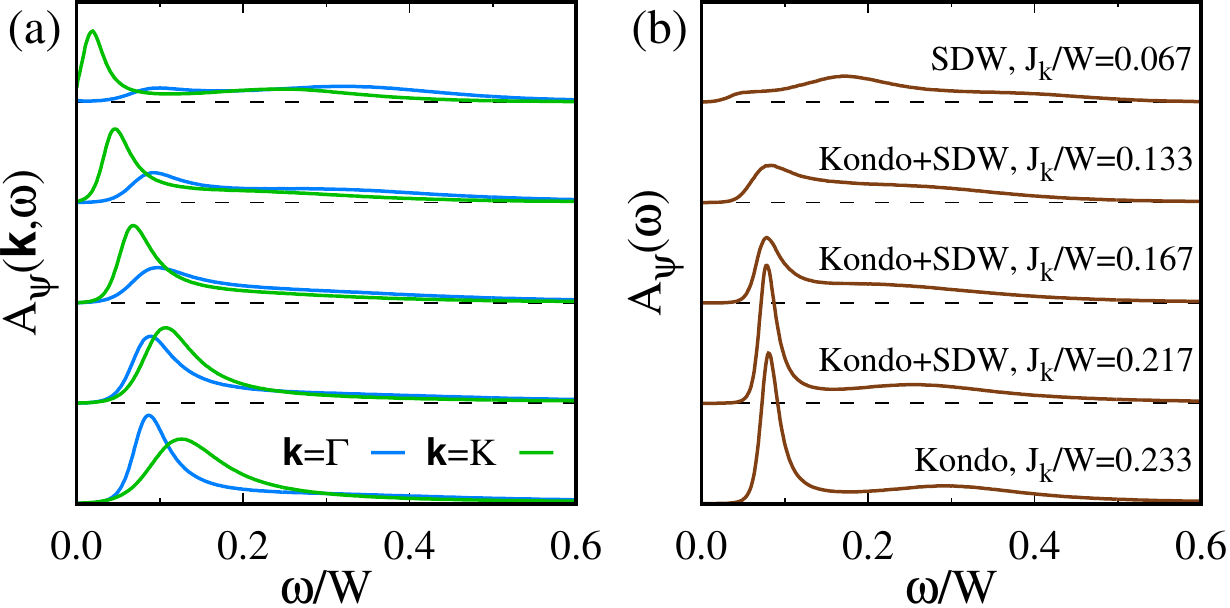}
	\caption{(a) Composite fermion spectral function $A_{\psi}(\ve{k},\omega)$  
	at the $\Gamma$ and Dirac $K$ points and (b) the corresponding density of states  
	$A_{\psi}(\omega)=\frac{1}{L^2}\sum_{\ve{k}}A_{\psi}(\ve{k},\omega)$  
	with decreasing (from bottom to top)  Kondo coupling $J_k$ obtained on the $L=18$ honeycomb KLM. 
}
\label{S:AJkL18}
\end{figure}

Figure~\ref{S:AJkL18}(a) illustrates the behavior of the composite fermion spectral function $A_{\psi}(\ve{k},\omega)$ 
at the $\Gamma$ and Dirac $K$ points for different values of Kondo coupling $J_k/W$.  
The corresponding density of states $A_{\psi}(\omega)=\frac{1}{L^2}\sum_{\ve{k}}A_{\psi}(\ve{k},\omega)$ is shown 
in Fig.~\ref{S:AJkL18}(b).

As can be seen, in the Kondo insulating phase at $J_k/W=0.233$, coherent Kondo screening  results in a well-defined peak at the 
$\Gamma$ momentum which determines the minimal  gap. The latter is also seen in $A_{\psi}(\omega)$ since the flat band of 
composite fermion quasiparticles generates a sharp peak that flanks the gap.   
The  low energy part of $A_{\psi}(\ve{k},\omega)$ evolves smoothly across the magnetic order-disorder transition 
at $J_k^c/W=0.2227(3)$ with a gradual shift of the minimal gap  from the $\Gamma$ point to the Dirac $K$ point. As is apparent, 
the change in the position of the minimal gap takes place away from $J_k^c$. Assuming a rigid band shift, the switch of Fermi wavevector 
in the metallic state at small dopings would lead to a change in the Fermi surface topology (Lifshitz transition).
Note however that this change in topology of the Fermi surface is unrelated to the breakdown of Kondo screening~\cite{Watanabe07}. 

As long as the magnetic order and low energy composite fermion band coexist (Kondo+SDW phase),  
one can track the signature of the quasiparticle band in $A_{\psi}(\omega)$. This should be contrasted with the SDW phase where in  
the absence of Kondo screening, signaled by a broad featureless spectrum at the $\Gamma$ point, an appropriate approach is 
the large-$S$ picture.  It accounts for the observed rearrangment of $A_{\psi}(\omega)$ such that a dominant contribution occurs 
at $\omega/W\simeq 1/6$.  It reflects the van Hove singularity in the conduction electron density of states.  
 
\subsection{Local spin-spin correlation function $S^{cf}$}

\begin{figure*}[h!]
\begin{center}
        \includegraphics[width=0.32\textwidth]{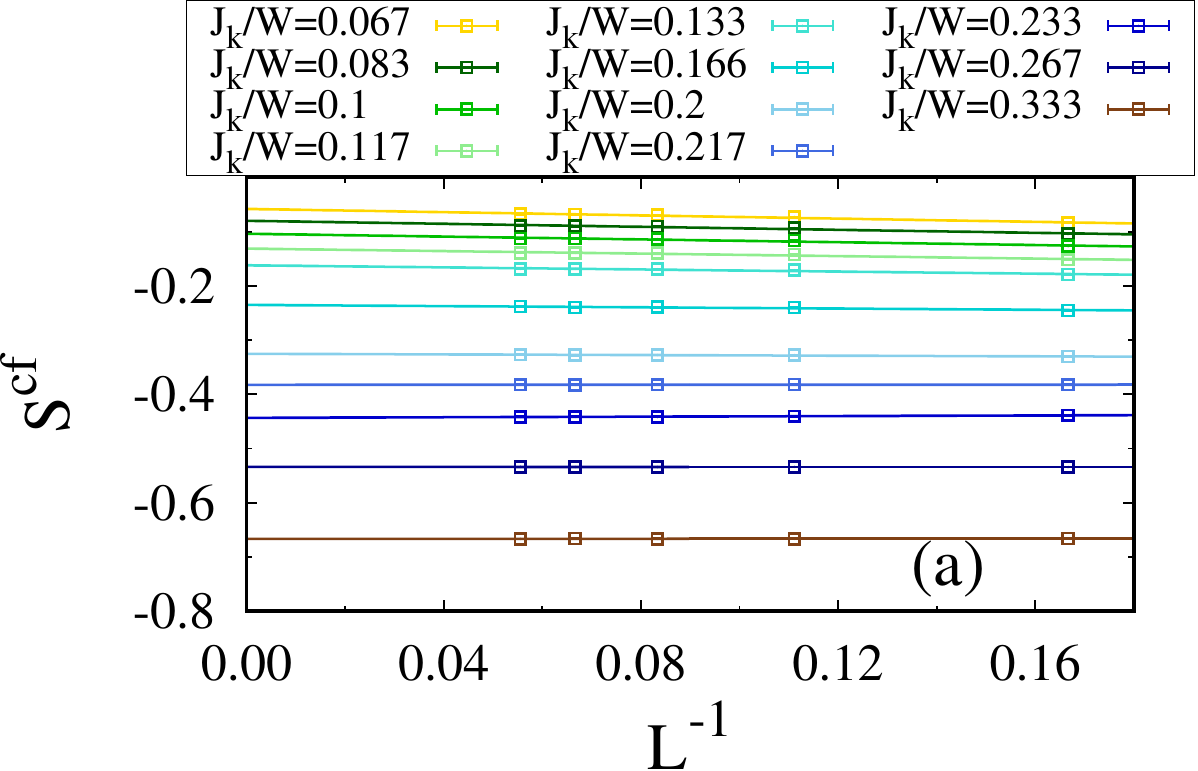}
        \includegraphics[width=0.32\textwidth]{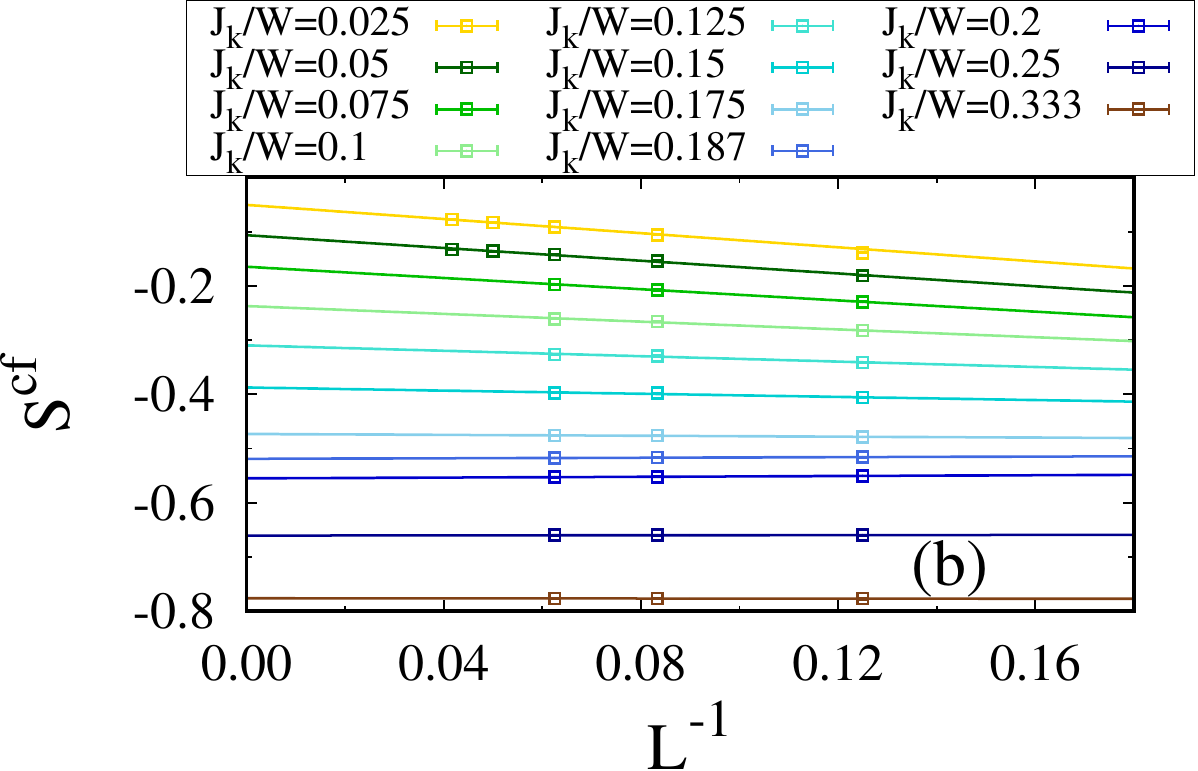}
\end{center}
\caption{Finite-size extrapolation of the local spin-spin correlation function $S^{cf}=\tfrac{2}{3N}\sum_{\ve{i}}\langle 
        \ve{\hat{c}}^{\dagger}_{\ve{i}} \ve{\sigma} \ve{\hat{c}}^{\phantom\dagger}_{\ve{i}} \cdot \ve{\hat{S}}^{}_{\ve{i}} \rangle$ 
	for the (a) honeycomb  and (b)  square KLMs. Solid lines are linear in $1/L$ fits to the QMC data. 
}
\label{S:Scf}
\end{figure*}

Figure~\ref{S:Scf} shows finite-size scaling of the local spin-spin correlation function 
\begin{equation}
S^{cf}=\frac{2}{3N}\sum_{\ve{i}}\langle
        \ve{\hat{c}}^{\dagger}_{\ve{i}} \ve{\sigma} \ve{\hat{c}}^{\phantom\dagger}_{\ve{i}} \cdot \ve{\hat{S}}^{}_{\ve{i}} \rangle
\end{equation}
which led us to the $J_k/W$-dependence of this quantity presented in the main text.

%% file: mean_field.tex
\input{intro-MF.tex}
\subsection{Large-$N$ mean-field approach}
The Kondo lattice Hamiltonian given in Eq.~(\ref{eqS:KLM}) in terms of the fermionic representation of spin operator 
$\hat{\S}_{\i}= \frac{1}{2} \sum_{\sigma,\sigma^\prime} \hat{f}^\dagger_{\i,\sigma} \ve{\sigma}_{\sigma,\sigma^\prime}  \hat{f}_{\i,\sigma^\prime}$  with the constraint $ \hat{f}^\dagger_{\i,\uparrow} \hat{f}_{\i,\uparrow}+\hat{f}^\dagger_{\i,\downarrow} \hat{f}_{\i,\downarrow} =1$   
can be written as,
\begin{eqnarray}
\hat{H}&=&\sum_{\k,\sigma} \epsilon_\k \hat{c}^\dagger_{\k,\sigma} \hat{c}_{\k,\sigma}
+\frac{J_k}{4} \sum_\i \big(\hat{f}^\dagger_{\i,\uparrow} \hat{f}_{\i,\uparrow} -\hat{f}^\dagger_{\i,\downarrow} \hat{f}_{\i,\downarrow}   \big) \big(\hat{c}^\dagger_{\i,\uparrow} \hat{c}_{\i,\uparrow} -\hat{c}^\dagger_{\i,\downarrow} \hat{c}_{\i,\downarrow} \big)
-\frac{J_k}{4}\sum_\i \big[\big(\hat{f}^\dagger_{\i,\uparrow} \hat{c}_{\i,\uparrow} +\hat{c}^\dagger_{\i,\downarrow} \hat{f}_{\i,\downarrow} \big)^2+\big(\hat{f}^\dagger_{\i,\downarrow} \hat{c}_{\i,\downarrow} +\hat{c}^\dagger_{\i,\uparrow} f_{\i,\uparrow} \big)^2 \big].\nonumber \\
\end{eqnarray}

In the large-$N$ approach we allow for Kondo screening as well as antiferromagnetic ordering with the following mean-field decouplings,
\begin{eqnarray}
\langle\hat{f}^\dagger_{\i,\uparrow} \hat{f}_{\i,\uparrow} -\hat{f}^\dagger_{\i,\downarrow} \hat{f}_{\i,\downarrow} \rangle=m_f e^{i \Q.\i},\quad 
\langle\hat{c}^\dagger_{\i,\uparrow} \hat{c}_{\i,\uparrow} -\hat{c}^\dagger_{\i,\downarrow} \hat{c}_{\i,\downarrow} \rangle=-m_c e^{i \Q.\i}, \quad 
\langle \hat{f}^\dagger_{\i,\uparrow} \hat{c}_{\i,\uparrow} +\hat{c}^\dagger_{\i,\downarrow} \hat{f}_{\i,\downarrow} \rangle =\langle\hat{f}^\dagger_{\i,\downarrow} \hat{c}_{\i,\downarrow} +\hat{c}^\dagger_{\i,\uparrow} f_{\i,\uparrow} \rangle=V.\nonumber \\
\end{eqnarray}
Here, $m_f$ denotes the staggered magnetization on localized spins,  $m_c$ denotes the staggered magnetization of conduction electrons, 
$V$ denotes the hybridization parameter between $\hat{c}$ and $\hat{f}$ electrons and $\Q$ is  the antiferromagnetic ordering wavevector.  %

For a honeycomb KLM  the mean-field Hamiltonian in the momentum space  can be written as follows,  
\begin{eqnarray}
\hat{H}_{mf}=\sum_{\k,\sigma} \phi^\dagger_{\k,\sigma} \left( \begin{array}{cccccccccccccccc}
\frac{J_k m_f\sigma}{4}&      Z(\k)&-\frac{J_kV}{2} & 0\\

~\\

Z^{\dagger}(\k)&      -\frac{J_k m_f \sigma}{4}& 0&-\frac{J_kV}{2} \\

~\\

 -\frac{J_k V}{2}& 0&  -\frac{J_km_c  \sigma}{4} & 0  \\

~\\

0 &  -\frac{J_k V}{2}& 0&   \frac{J_k m_c \sigma }{4}  
\end{array}\right)\hat{\phi}_{\k,\sigma}+e_0 N_u.\nonumber \\
\label{mf_ham_HKLM}
\end{eqnarray}
Here,  $\hat{\phi}^\dagger_{\k,\sigma}=\big\{\hat{\phi}^\dagger_{\hat{c}_{\k,a},\sigma}, \hat{\phi}^\dagger_{\hat{c}_{\k,b},\sigma},  \hat{\phi}^\dagger_{\hat{f}_{\k,a},\sigma} ,  \hat{\phi}^\dagger_{\hat{f}_{\k,b},\sigma}\big\}$, $e_0=\big( \frac{J_k V^2}{2}+\frac{J_k m_fm_c}{4} \big)$,  $Z(\k)=-t(1+e^{-i \k\cdot\ve{a}_2} +e^{-i \k\cdot(\ve{a}_2-\ve{a}_1)})$ with $\ve{a}_1=(1,0)$ and $\ve{a}_2=(\frac{1}{2},\frac {\sqrt{3}} {2})$ and $N_u$ is the number of unit cells.

Diagonalization of the mean-field  Hamiltonian gives the following dispersion relations,
\begin{eqnarray}
E_{\k,n}&=&\mp\frac{1}{4} \sqrt{P_\k \mp\frac{1}{2} \sqrt {Q_\k- 4 R_\k}}
\label{Enl_vs_k_HKLM}
\end{eqnarray}
with,
%
\begin{align}
	P_\k&=8 |Z(\k)|^2 + (m_c^2 J^2_k)/2 + (m_f^2 J^2_k)/2+ 4 J^2_k V^2, \nonumber \\ 
	Q_\k&=\big(-16 |Z(\k)|^2 - m_c^2 J^2_k - m_f^2 J^2_k- 8 J^2_k V^2\big)^2,\nonumber\\
	R_\k&=\big(16 m_c^2 |Z(\k)|^2 J^2_k + m_c^2 m_f^2 J^4_k + 8 m_c m_f J^4_k V^2+ 16 J^4_k V^4\big)\nonumber.
\end{align}

 The  ground state energy per unit cell can be computed as, 
\begin{eqnarray}
 e_g=e_0+\frac{2}{N_u} \sum_{\k, n, E_{\k,n}<0 } E_{\k,n}.
\end{eqnarray}
The self consistent equations of  mean-field parameters can obtained from the saddle point approximation,
\begin{eqnarray}
\frac{d e_g}{d V}=0, \quad \frac{d e_g}{d m_f}=0,  \quad \frac{d e_g}{d m_c}=0.
\end{eqnarray}

  \begin{figure}[h]
\centering
\includegraphics[width=0.45\textwidth]{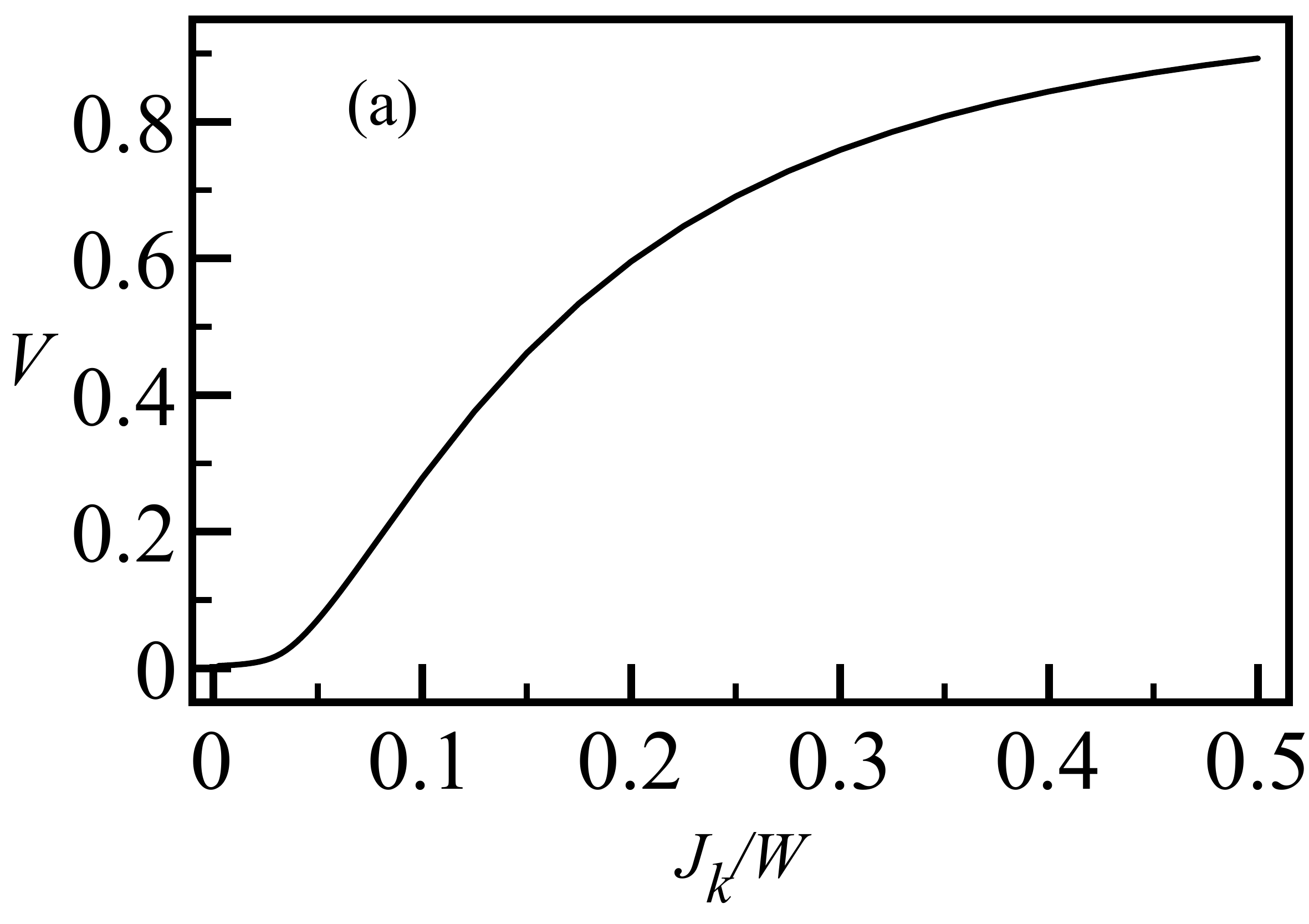} 
\includegraphics[width=0.45\textwidth]{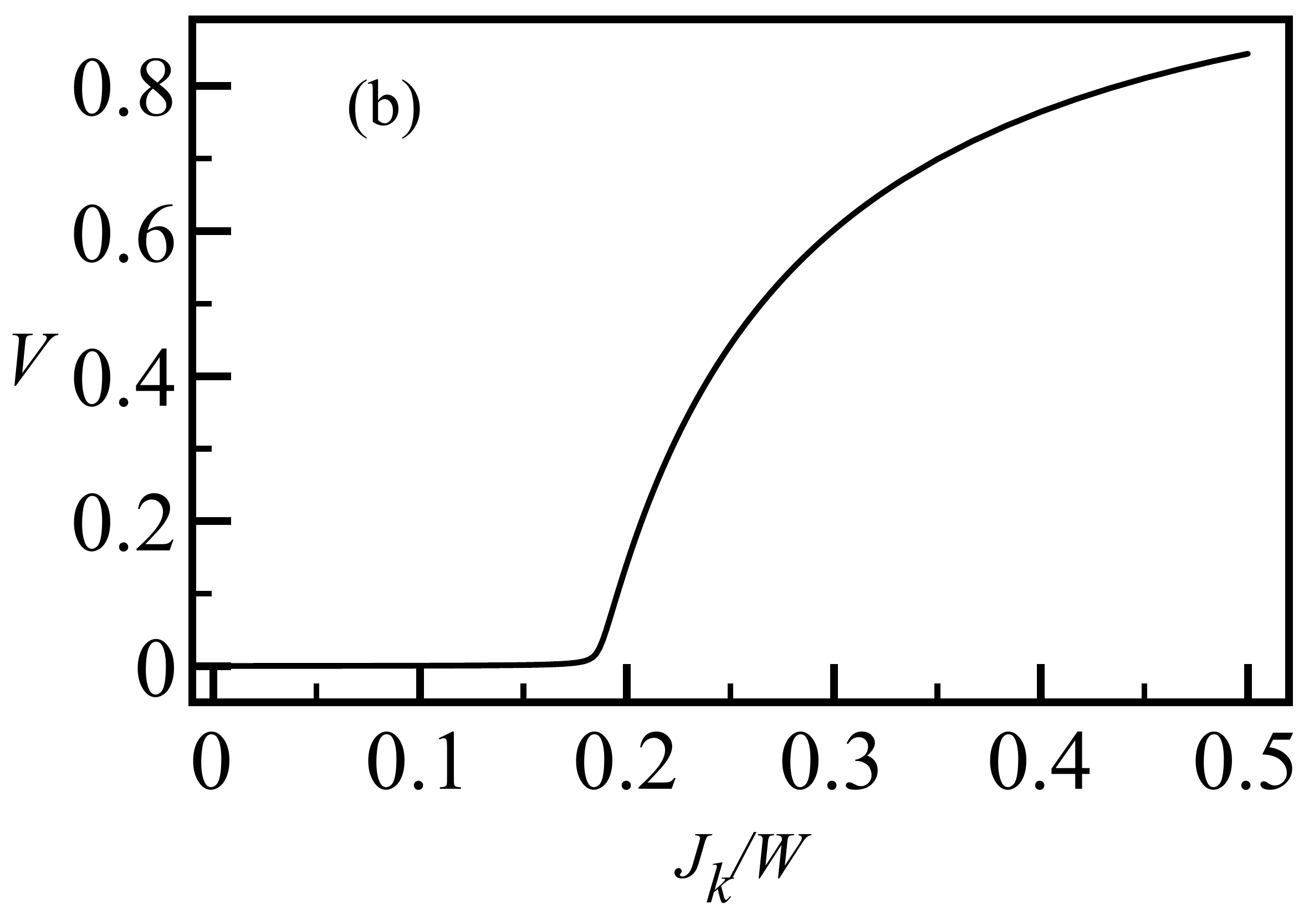} 
\caption{Hybridization parameter $V$ as a function of $J_k/W$  in the absence of magnetic order computed 
	  within the large-$N$ mean-field approach.  (a) For a square KLM.  (b) For a honeycomb KLM. }  
\label{largeV_vsJk} 
\end{figure}

  \begin{figure}[h]
\centering
\includegraphics[width=0.55\textwidth]{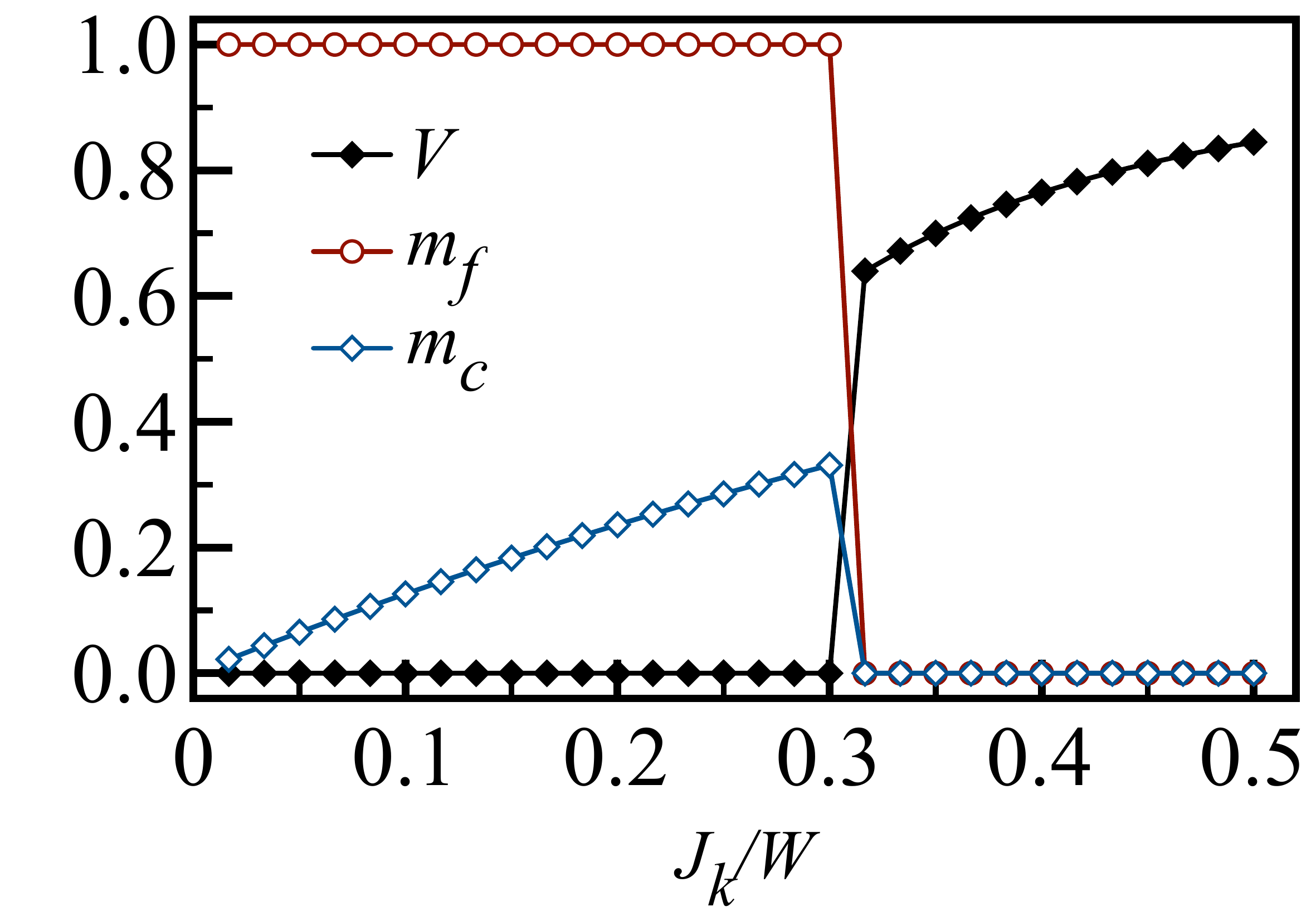} 
\caption{Mean-field  order parameters as a function of  $J_k/W$ for the honeycomb KLM within the large-$N$ mean-field approach.}  
\label{largeNmV_vsJk} 
\end{figure}

Figures~\ref{largeV_vsJk}(a) and \ref{largeV_vsJk}(b) plot  the hybridization order parameter as a function of 
$J_k/W$  on the square and honeycomb lattices in the absence of magnetic order. 
Figure~\ref{largeNmV_vsJk} was obtained by taking into account both Kondo screening and magnetic order   and 
plots the  mean-field order parameters  $m_f$, $m_c$, and $V$ as a function of $J_k/W$ for the honeycomb KLM.

\subsection {Bond fermion mean-field theory}
\label{Sec:bond_mean_field}
To formulate the bond fermion mean-field theory we consider the states:  
\begin{eqnarray}
	   & & \hat{s}^{\dagger}_{\ve{i}}\big|0\big\rangle  =  \frac{1}{\sqrt{2}}   \left(\hat{c}^{\dagger}_{\ve{i}, \uparrow}  \hat{f}^{\dagger}_{\ve{i}, \downarrow}   
	    - \hat{c}^{\dagger}_{\ve{i}, \downarrow} \hat{f}^{\dagger}_{\ve{i}, \uparrow}   \right)\big|0\big\rangle     \nonumber  \\ 
	   & &\hat{t}^{\dagger}_{\ve{i},0}\big|0\big\rangle  =  \frac{1}{\sqrt{2}}   \left(\hat{c}^{\dagger}_{\ve{i}, \uparrow} \hat{f}^{\dagger}_{\ve{i}, \downarrow} 
	    +   \hat{c}^{\dagger}_{\ve{i}, \downarrow}   \hat{f}^{\dagger}_{\ve{i}, \uparrow}   \right)\big|0\big\rangle \nonumber    \\
	   & & \hat{t}^{\dagger}_{\ve{i},\sigma}\big|0\big\rangle  =  \hat{c}^{\dagger}_{\ve{i}, \sigma} \hat{f}^{\dagger}_{\ve{i}, \sigma}\big|0\big\rangle   \nonumber    \\
	  & & 
	     \hat{h}^{\dagger}_{\ve{i},\sigma}\big|0\big\rangle = \hat{f}^{\dagger}_{\ve{i},\sigma} \big|0\big\rangle   \nonumber    \\  & & 
	      \hat{d}^{\dagger}_{\ve{i},\sigma}\big|0\big\rangle =\hat{c}^{\dagger}_{\ve{i}, \uparrow} \hat{c}^{\dagger}_{\ve{i}, \downarrow}  \hat{f}^{\dagger}_{\ve{i},\sigma} \big|0\big\rangle.
\end{eqnarray}
Here, $\hat{s}^{\dagger}$ and  $\hat{t}^{\dagger}_{1,0,-1}$ denote a singlet and three triplet states  with one conduction electron per site  and  $\hat{h}^{\dagger}_{\sigma}$ and  $\hat{d}^{\dagger}_{\sigma}$ denote holons  and doublons  of the conduction electrons.   
In this representation, the constraint
 \begin{eqnarray}
\hat{s}^\dagger_{\ve{i}}\hat{s}_{\ve{i}}+\sum_{m=1,0,-1}\hat{t}^\dagger_{\ve{i},m}\hat{t}_{\ve{i},m}+\sum_{\sigma=\uparrow,\downarrow}(\hat{h}^\dagger_{\ve{i},\sigma} \hat{h}_{\ve{i},\sigma}+\hat{d}^\dagger_{\ve{i},\sigma} \hat{d}_{\ve{i},\sigma})=1
\end{eqnarray} 
suppresses the unphysical  states.

The conduction electron  operator and the spin operators on $\hat {f}$ and $\hat{c}$ electrons in the above representation take the 
 following forms,
\begin{eqnarray}
\label{c_bond}
\hat{c}^\dagger_{\i,\sigma}=\frac{\sigma}{\sqrt{2}} (\hat{s}^\dagger_{\ve{i}}+\sigma \hat{t}^\dagger_{\ve{i},0}) \hat{h}_{\ve{i},-\sigma}+\hat{t}^\dagger_{\ve{i},\sigma} \hat{h}_{\ve{i},\sigma}-\frac{\hat{d}^\dagger_{\ve{i},\sigma}}{\sqrt{2}} (\hat{s}_{\ve{i}}-\sigma \hat{t}_{\ve{i},0}) +\sigma \hat{d}^\dagger_{\ve{i},-\sigma} \hat{t} _{\ve{i},-\sigma}\\
 \hat{S}_{\i,\alpha}=\frac{1}{2} \big(\hat{s}^\dagger_{\i} \hat{t}_{\i,\alpha}+ \hat{t}^\dagger_{\i,\alpha} \hat{s}_{\i}-i\epsilon_{\alpha\beta\gamma} \hat{t}^\dagger_{\i,\beta} \hat{t}_{\i,\gamma}  \big)+\hat{S}^h_{\i,\alpha}+ \hat{S}^d_{\i,\alpha}\\
 \hat{S}^c_{\i,\alpha}=\frac{1}{2} \big(-\hat{s}^\dagger_{\i} \hat{t}_{\i,\alpha}- \hat{t}^\dagger_{\i,\alpha} \hat{s}_{\i}-i\epsilon_{\alpha\beta\gamma} \hat{t}^\dagger_{\i,\beta} \hat{t}_{\i,\gamma}  \big)
\end{eqnarray}
where $\epsilon_{\alpha\beta\gamma}$ is the totally antisymmetric tensor, the holon and doublon spin operators read  
$\hat{S}^h_{\i,\alpha}= \frac{1}{2} \sum_{\sigma,\sigma^\prime} \hat{h}^\dagger_{\i,\sigma} \ve{\sigma}_{\sigma,\sigma^\prime}  \hat{h}_{\i,\sigma^\prime}$   and 
$\hat{S}^d_{\i,\alpha}= \frac{1}{2} \sum_{\sigma,\sigma^\prime} \hat{d}^\dagger_{\i,\sigma} \ve{\sigma}_{\sigma,\sigma^\prime}  \hat{d}_{\i,\sigma^\prime}$ and  the triplon operators are defined as follows,
 \begin{eqnarray}
\hat{t}^{\dagger}_{\ve{i},z}\big|0\big\rangle  = \hat{t}^{\dagger}_{\ve{i},0}\big|0\big\rangle, \quad \hat{t}^{\dagger}_{\ve{i},x}\big|0\big\rangle  =\frac{1}{\sqrt{2}}\Big( \hat{t}^{\dagger}_{\ve{i},1}+\hat{t}^{\dagger}_{\ve{i},-1} \Big)\big|0\big\rangle, \quad \hat{t}^{\dagger}_{\ve{i},y}\big|0\big\rangle  =-\frac{i}{\sqrt{2}}\Big( \hat{t}^{\dagger}_{\ve{i},1}-\hat{t}^{\dagger}_{\ve{i},-1} \Big)\big|0\big\rangle. 
\end{eqnarray}

In the bond fermion mean-field approach to the KLM, the Kondo phase corresponds to condensation of singlets $\langle \hat{s}_\i\rangle$  
and the SDW phase corresponds to condensation of $z$ component of triplets $\langle \hat{t}_{\i,z}\rangle$ in the  singlet background. Hence, we consider the following mean-field approximation~\cite{PhysRevB.64.092406}: 

 \begin{eqnarray}
 \langle \hat{s}_\i\rangle=\langle\hat{s}^\dagger_\i\ \rangle=s\\
  \langle \hat{t}_{\i,z}\rangle=\langle\hat{t}^\dagger_{\i,z}\ \rangle=(-1)^{\i}m
 \end{eqnarray} 
where $(-1)^\i = 1 (-1) $ on  sub-lattice  A (B), corresponds  to  antiferromagnetic  ordering.

 Using the above approximation  the honeycomb Kondo lattice mean-field Hamiltonian can be written as,
 \begin{eqnarray}
  \hat{H}_{mf}&= &e_0N_u+\mu \sum_{\i,\sigma} \Big(\hat{h}^{\dagger a}_{\ve{i},\sigma} \hat{h}^a_{\ve{i},\sigma}-\hat{d}^a_{\ve{i},\sigma} \hat{d}^{^\dagger a}_{\ve{i},\sigma}+\hat{h}^{\dagger b}_{\ve{i},\sigma} \hat{h}^b_{\ve{i},\sigma}-\hat{d}^b_{\ve{i},\sigma} \hat{d}^{^\dagger b}_{\ve{i},\sigma} \Big)+\frac {1} {2}(s^2-m^2) \sum_{\langle \i \j\rangle,\sigma} \Big( \hat{h}^{\dagger a}_{\i,\sigma} \hat{h}^b_{\j,\sigma}+ \hat{d}^{ a}_{\i,\sigma} \hat{d}^{ \dagger b}_{\j,\sigma}+ \text{H.c} \Big)\nonumber\\  &&
 + \frac {1} {2}(s+m)^2 \sum_{\langle \i \j\rangle,\sigma} \Big( -\hat{h}^{\dagger a}_{\i,-\sigma} \hat{d}^{\dagger b}_{\j,\sigma}+ \hat{d}^{ a}_{\i,-\sigma} \hat{h}^{ b}_{\j,\sigma}+ \text{H.c} \Big) + \frac {1} {2}(s-m)^2 \sum_{\langle \i \j\rangle,\sigma} \Big( \hat{h}^{\dagger a}_{\i,\sigma} \hat{d}^{ \dagger b}_{\j,-\sigma} -\hat{d}^{a}_{\i,\sigma} \hat{h}^{b}_{\j,\sigma}+ \text{H.c} \Big).
 \end{eqnarray} 
Here, $e_0=\big( -\frac{3 J_k}{4} s^2+\frac{J_k}{4}m^2+\mu(s^2+m^2+1)\big)$ and $N_u$ is the number  of unit cells. We use the global 
Lagrange multiplier $\mu_\i=\mu$  to impose the constraint $\mu (s^2+m^2-1)$.  Note that in the above we have ignored all  the terms corresponding to 
the transverse  and longitudinal spin fluctuations.

The mean-field Hamiltonian in momentum  space can be written as follows,
 \begin{eqnarray}
  \hat{H}_{mf}&= &e_0N_u+\sum_{\k,\sigma} \phi^\dagger_{\k ,\sigma} M(\k) \phi_{\k,\sigma}
 \end{eqnarray}   
where  $\phi^\dagger_{\k,\sigma}=\{\hat{h}^{\dagger a}_{\k,\sigma},\hat{h}^{\dagger b}_{\k,\sigma},\hat{d}^{\dagger a}_{-\k,\sigma},\hat{d}^{\dagger b}_{-\k,\sigma}\}$  
and the  matrix  $M(\k)$
\begin{eqnarray*}
 M(\k) =\left( \begin{array}{cccccccccccccccc}
\mu~~&\alpha_\k~~&0~~&0~~&0~~&0~~&0~~&\beta_k \\ \\

\alpha^\dagger_\k~~&\mu~~&0~~&0~~&0~~&0~~&\gamma^\dagger_\k~~&0 \\ \\

~

0~~&0~~&\mu~~&\alpha_\k~~&0~~&-\gamma_\k~~&0~~&0 \\ \\

~

0~~&0~~&\alpha^\dagger_\k~~&\mu~~&-\beta^\dagger_\k~~&0~~&0 ~~&0 \\ \\

~

0~~&0~~&0~~&-\beta_\k~~&-\mu~~&\alpha^\dagger_\k~~&0~~&0 \\ \\

~

0~~&0~~&-\gamma^\dagger_\k~~&0~~&\alpha_\k~~&-\mu~~&0~~&0 \\ \\

~

0~~&\gamma_\k~~&0~~&0~~&0~~&0~~&-\mu~~&\alpha^\dagger_\k \\ \\

\beta^\dagger_\k~~&0~~&0~~&0~~&0~~&0~~&\alpha_\k~~&-\mu \\
\end{array}
\right)
\end{eqnarray*}
 with
 \begin{eqnarray}
 \alpha_\k=- \frac {1} {2}(s^2-m^2) Z(\k), \quad  \beta_\k=- \frac {1} {2}(s+m)^2 Z(\k), \quad  \gamma_\k=- \frac {1} {2}(s-m)^2  Z(\k), \quad  Z(\k)=-t(1+e^{-i \k\cdot\ve{a}_2} +e^{-i \k\cdot(\ve{a}_2-\ve{a}_1)}).\nonumber
    \end{eqnarray}

The matrix $M(\k)$  can be diagonalized via unitary transformation which gives the following dispersion relations for $z$ component of 
spin $\sigma=\uparrow(\downarrow)$, 
\begin{eqnarray}
E_{\k,n}=\mp\sqrt{\frac{1}{2} (m^2+s^2)^2 | Z(\k)|^2+\mu^2\mp2\sqrt{\frac{1}{4} \mu^2 | Z(\k)|^2 (m^2-s^2)^2+\frac{1}{16} | Z(\k)|^4 (m^2+s^2)^4 }}.
\end{eqnarray}

The mean-field ground state energy per unit cell can be computed as follows,
\begin{eqnarray}
 e_g=e_0+\frac{2}{N_u} \sum_{\k,n, E_{\k,n}<0 } E_{\k,n}.
 \end{eqnarray}

Next, using the saddle point approximation,
     \begin{eqnarray} 
     \frac{d e_g}{ d s}=0, \quad      \frac{d e_g}{ d \mu}=0, \quad      \frac{d e_g}{ d m}=0
       \end{eqnarray}
we obtain the following self consistent equations for mean-field parameters $\mu$, $m^2$, and $s^2$,
 \begin{eqnarray} 
 \mu=\frac{J_k}{4}-\frac{1}{2N_u} \sum_\k  |Z(\k)|^2 (m^2+s^2) \Big( \frac{1}{E_{\k,1}}+ \frac{1}{E_{\k,2}}\Big)-\frac{1}{4N_u} \sum_\k  \frac{ |Z(\k)|^4 (m^2+s^2)^3  }{2A_\k}      \Big(\frac{1}{E_{\k,1}}- \frac{1}{E_{\k,2}}\Big) \\
 {m}^2=\frac{1}{2N_u} \sum_\k  \frac{|Z(\k)|^2 \mu^2{m}^2 (s^2-m^2) }{J_kA_\k}      \Big(\frac{1}{E_{\k,1}}-\frac{1}{E_{\k,2}}\Big) \\
 s^2=-1- m^2-\frac{1}{N_u} \sum_\k \mu \Big( \frac{1}{E_{\k,1}}+ \frac{1}{E_{\k,2}}\Big)-\frac{1}{N_u} \sum_\k \frac {|Z(\k)|^2 (s^2-m^2)^2 \mu}{4A_\k}  \Big(\frac{1}{E_{\k,1}}- \frac{1}{E_{\k,2}}\Big)
\end{eqnarray}
where ${E_{\k,1}}$ and ${E_{\k,2}}$ are the two lowest quasiparticle bands  and $A_\k$ has the following form,
   \begin{eqnarray}
 A_\k=\sqrt{\frac{1}{4} \mu^2 | Z(\k)|^2 (m^2-s^2)^2+\frac{1}{16} | Z(\k)|^4 (m^2+s^2)^4 }.\nonumber
  \end{eqnarray}
  
  The magnetization of  $c$ and $f$ electrons can be computed as  follows,
 \begin{eqnarray}
 m_c=\frac{2}{N_u} \sum_\i (-1)^\i \langle \hat{S}^c_{z,\i} \rangle=2 m{s} \\
 m_f=\frac{2}{N_u} \sum_\i (-1)^\i \langle \hat{S}_{z,\i} \rangle=2 m {s} +\frac{1}{N_u} \sum_\k \frac{ 2 |Z(\k)|^2 \mu m s (s^2+m^2)}{E_{1,\k} E_{2,\k} (E_{1,\k}+E_{2,\k})}.
 \end{eqnarray}

\begin{figure}[htbp]
\centering
\includegraphics[width=0.55\textwidth]{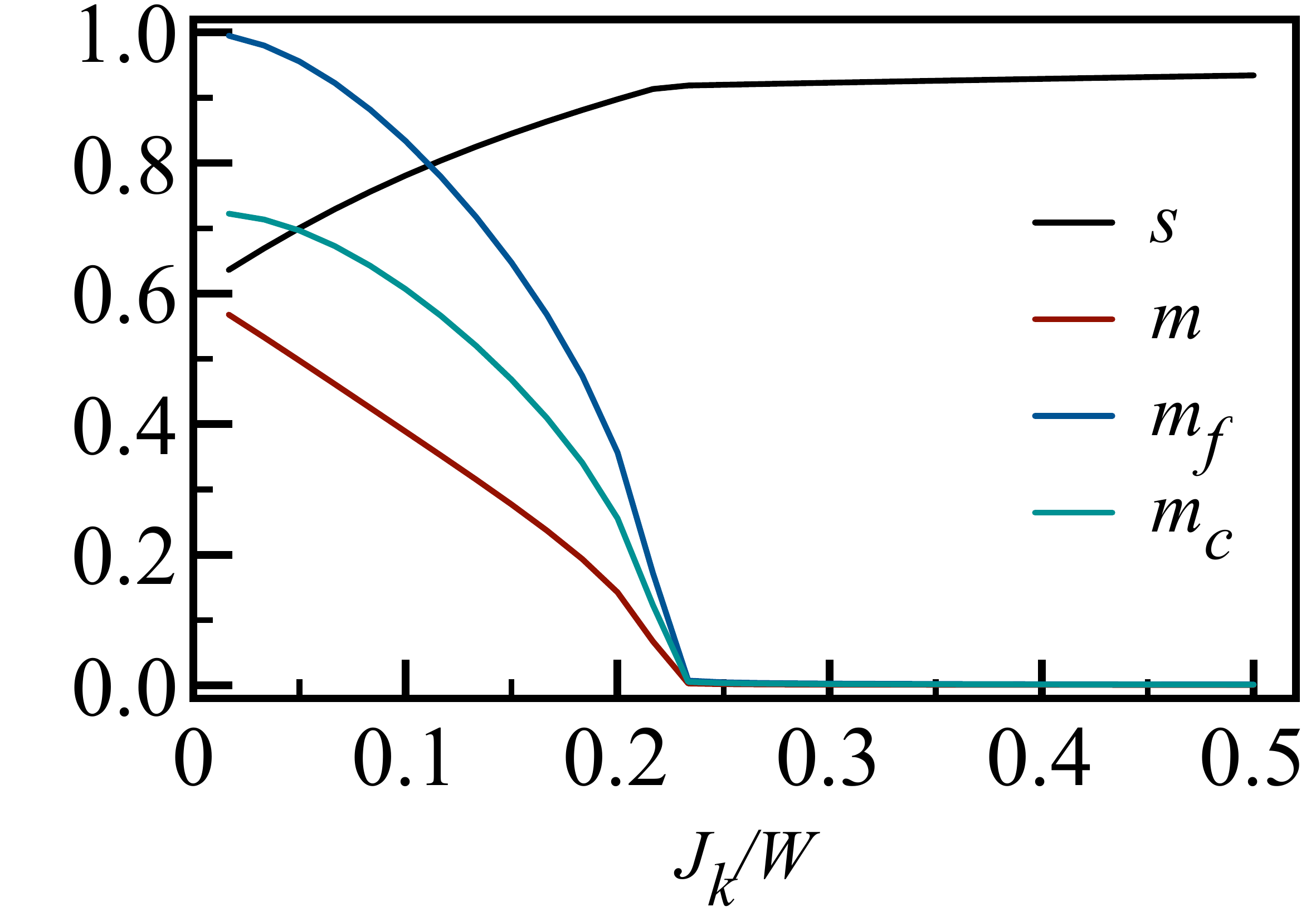} 
\caption{Mean-field  order parameters as a function of  $J_k/W$ for the honeycomb KLM  within the bond fermion mean-field approach.  }  
\label{bondfermion_mf} 
\end{figure}

 Figure~\ref{bondfermion_mf}  plots the  mean-field order parameters obtained within the bond fermion mean-field approach as a function of $J_k/W$ for the honeycomb KLM.

%% file: intro-MF.tex
\section{Mean-field approximations} 
In this  section,  we  review  mean-field  approximations    in the  aim  of  providing  an  account of  our  results.  
We  will see  that  both  the  large-$N$ as  well as  the bond fermion mean-field  approximations  fail  at providing  \textit{consistent}  
results  for  the  square  and  honeycomb lattices.  

In the large-$N$  mean-field approximation    we   neglect   the  fluctuations of  the  boson  field  $b_{\ve{i}}(\tau)$    and  
take  into  account  the  constraint  on average.    The field $b_{\ve{i}}(\tau)$   possesses   phase  as  well as  amplitude  
fluctuations.  Phase  fluctuations  will not  be taken into account   at the  mean-field  level  and  the only  manner in 
which     Kondo breakdown  can occur  is  through the  vanishing of  the amplitude  of    the  boson  field.     
This  actually  stands   at odds  with   our  QMC  data  that   suggest  that the  amplitude of  the field  remains  constant  
for  all values of  the Kondo coupling and that  Kondo  breakdown  occurs  due  to phase  fluctuations.    In the  single  impurity limit,     
or  equivalently in the  absence  of   magnetic  ordering,  this approach  does  capture  the  differences  between the  honeycomb 
and  square lattices, see Fig.~\ref{largeV_vsJk}.      However,  when  magnetic  ordering, alongside  Kondo  screening  is  included,   
the  approximation yields \textit{ Kondo breakdown}  in the magnetic phase  for  \textit{both}  the honeycomb (see Fig.~\ref{largeNmV_vsJk} 
and also Ref.~\cite{EPJB.86.195})   and  square ~\cite{PhysRevB.62.76,Capponi00}  lattices.    

An  approximation  that    captures  the coexistence  of  Kondo  screening  and magnetism on the square lattice,   is  the  bond fermion  
mean-field  approximation~\cite{PhysRevB.64.092406,PhysRevB.98.245125}.    In this  strong  coupling  approach, the  Kondo effect is 
accounted  for    by   a     vacuum  expectation  value  of  the  the  singlet  correlator,  
$\hat{s}^{\dagger}_{\ve{i}} =  \frac{1}{\sqrt{2}}   \left(\hat{c}^{\dagger}_{\ve{i}, \uparrow}  \hat{f}^{\dagger}_{\ve{i}, \downarrow}  - \hat{c}^{\dagger}_{\ve{i}, \downarrow}  \hat{f}^{\dagger}_{\ve{i}, \uparrow}   \right) $.      Since   $  \left< | b_{\ve{i}} |^2  \right>   \propto   
\left<   \hat{V}^{\dagger}_{\ve{i}}  \hat{V}^{}_{\ve{i}}  \right>    \propto  \left<   \frac{1}{2} \hat{\ve{c}}^{\dagger}_{\ve{i}}  \ve{\sigma}   \hat{\ve{c}}^{}_{\ve{i}}  \cdot  \hat{\ve{S}}_{\ve{i}}  \right>  
\propto  \left< \hat{s}^{\dagger}_{\ve{i}}  \hat{s}^{}_{\ve{i}} \right>       $,   a  non-vanishing  vacuum expectation value  
of   $\hat{s}^{\dagger}_{\ve{i}} $, $s$,    corresponds  to the  Kondo effect.  By  virtue  of  completeness,  we  have  included  
this  approximation   in  subsection  \ref{Sec:bond_mean_field},  and  the  reader  will  convince  oneself  that  only  solutions 
 with  a  finite  value of  $s$ will  occur, see Fig.~\ref{bondfermion_mf}.  As  such   this  approximation  invariably  predicts   
\textit{coexistence} of  magnetism  and   the Kondo effect,  for  \textit{both} the  square  and  honeycomb lattices.    
This  again stands  at  odds  with our QMC  results of the main text. 